\documentclass[final,3p,times,twocolumn]{elsarticle}
\usepackage[utf8]{inputenc}
\usepackage[T1]{fontenc}
\usepackage{algorithmic}
\usepackage{array}
\usepackage{textcomp}
\usepackage{stfloats}
\usepackage{url}
\usepackage{verbatim}
\usepackage{graphicx}
\usepackage{balance}
\usepackage{booktabs}
\usepackage{lipsum}
\usepackage{caption}
\usepackage{amsmath,amssymb,amsfonts,bm}
\usepackage{algorithmic}
\usepackage{graphicx}
\usepackage{textcomp}
\usepackage{stfloats}
\usepackage{placeins}
\usepackage[ruled,vlined]{algorithm2e}
\usepackage[table,xcdraw]{xcolor}
\usepackage{multirow}
\usepackage{appendix}
\usepackage{xurl}

\usepackage{ragged2e}

\newlength{\ProblemColumnWidth}
\newlength{\IndustrialMeaningColumnWidth}
\newlength{\ConsequenceColumnWidth}
\newlength{\ChasmRelationColumnWidth}

\newcolumntype{P}[1]{>{\RaggedRight\arraybackslash}p{#1}}

\begin{document}

\begin{frontmatter}

\title{Unsupervised and explainable authorization of change point in multivariate industrial data for trustworthy system and process monitoring}

 \author[label1]{Justyna Witulska}\corref{cor1}\ead{justyna.witulska@pwr.edu.pl}
 \author[label2]{Tomasz Barszcz}
  \author[label1]{Agnieszka Wy\l oma\'nska}
 \author[label3,label4]{Ireneusz Jab\l o\'nski} 
 \affiliation[label1]{organization={Faculty of Pure and Applied Mathematics, Hugo Steinhaus Center, Wroclaw University of Science and Technology},
             addressline={Wyspianskiego 27}, 
             city={Wrocław},
             postcode={50-370}, 
             country={Poland}}

 \affiliation[label2]{organization={Faculty of Mechanical Engineering and Robotics, AGH University},
             addressline={al. Mickiewicza 30}, 
             city={Kraków},
             postcode={30-059}, 
             country={Poland}}

 \affiliation[label3]{organization={Faculty MINT, Brandenburg University of Technology},
             addressline={Konrad-Zuse-Straße 1}, 
             city={Cottbus},
             postcode={03046}, 
             country={Germany}}

 \affiliation[label4]{organization={Fraunhofer Institute for Photonic Microsystems},
 addressline={Maria-Reiche-Straße 2}, 
 city={Dresden},
 postcode={01109}, 
 country={Germany}}

\begin{abstract}
The paper introduces a Change point Hypothesis Assessment for Significance in Multivariate data (CHASM) -- a novel non-parametric approach for evaluating the significance of change points detected in multivariate data, especially relevant for a trustworthy autonomous industrial system. The method is highly scalable, doesn't require data labeling, and allows for customizable evaluation criteria, such as inter-component relationships. Its non-parametric nature broadens applicability to non-Gaussian distributed data. CHASM incorporates robust auxiliary statistics that maintain reliability even in the presence of outliers, enabling accurate assessment of change point significance for complex data. Beyond statistical significance testing, CHASM also facilitates root-cause analysis of detected changes. It offers a solution expected by practitioners for validating change points identified by segmentation-based algorithms, particularly those lacking rigorous statistical foundations. To support situation aware decision making, a {trustworthy factor} is introduced and its performance demonstrated for simulated and real-world data recorded in machine monitoring system. Using MAPE, it was evidenced that the two-stage procedure –  identification of the change point and authorization of the change point – significantly outperforms (23-fold in the reported experiment) the one-stage scheme. All these are essential in real-world applications, where overfitting and false alarms can lead to loss of trust in technical solutions.
\end{abstract}

\begin{keyword}
multivariate analysis \sep non-Gaussian distribution \sep root cause analysis \sep significance testing methods \sep segmentation \sep industrial data
\end{keyword}

\end{frontmatter}

\begin{figure*}[h!]
    \centering
    \includegraphics[width=0.8\linewidth]{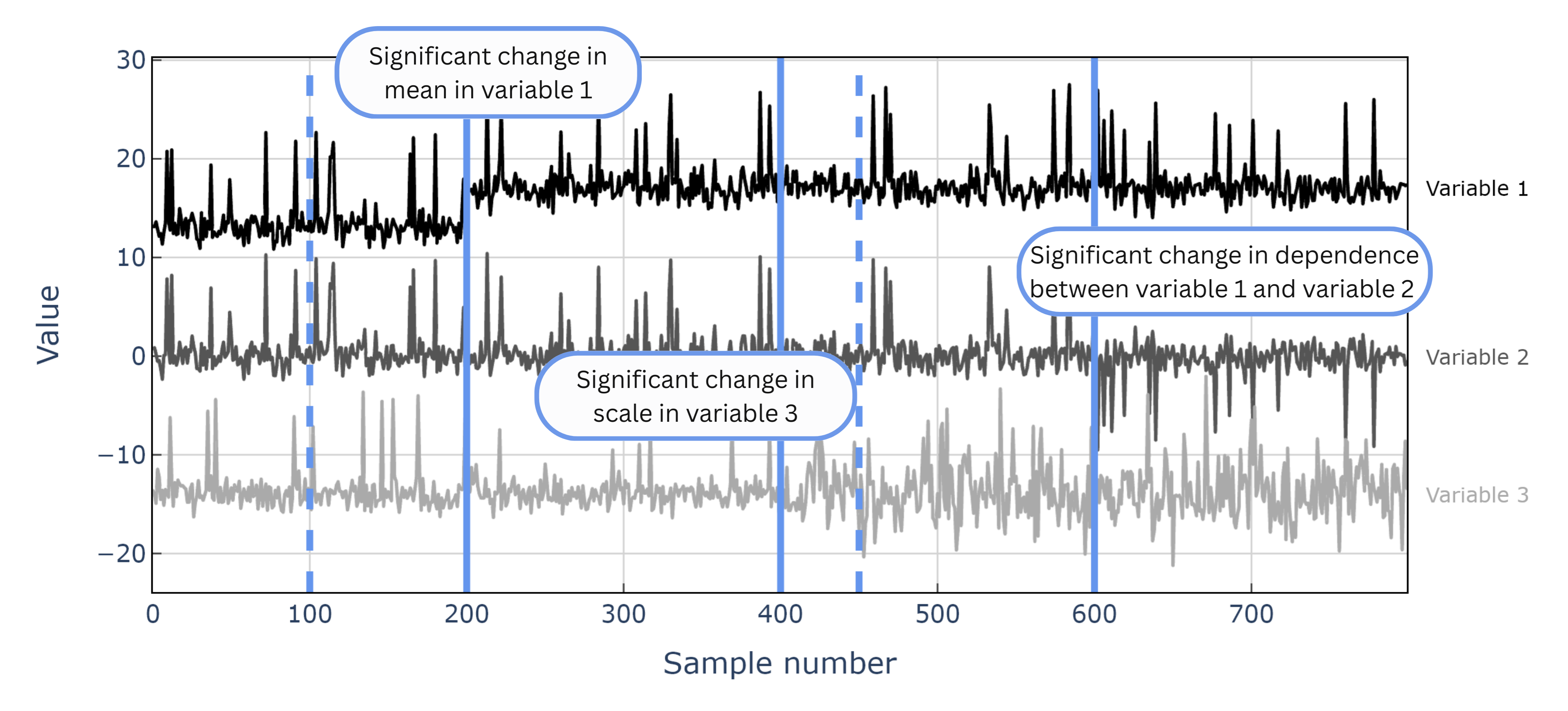}
    \caption{An example of three-dimensional contaminated data for which we can observe distributional changes over time: (a) change in mean for variable 1 (at 200 sample), (b) change in scale of the variable 3 (at 400 sample) and (c) change in dependence between variable 1 and variable 2 (at 600 sample). Blue lines correspond to change points detected by basic segmentation algorithm (dashed lines correspond to false positives and solid lines are true change points for which we can observe significant distributional change).} 
    \label{fig:idea_intro}
\end{figure*}

\section{Introduction}
\label{sec:introduction}
\subsection{Problem description and motivation}
The industrial realization of the concept of autonomy, particularly relevant to complex infrastructures and processes, is heavily conditioned by (multi)sensor and data-driven capabilities. These resources have also initiated a migration from manual and subjective inspection performed in this infrastructure by domain experts toward semi-supervised (human-in-the-loop) solutions, in which the identification of following states in an evolving system and its accompanying processes is algorithmized, and its operation is temporarily verified and authorized \cite{CPD_Significance_Industrial_1, CPD_Multisenor_WSN_challenge, CPD_Review_GUPTA_ESA, critical_proc_syst_automation, Situational_Awareness_3}. Both the patterns in the data as well as the knowledge and experience represented (and used explanatively/interpretively) by the expert intermingle, determining the authorization decision in terms of maximizing the probability that the observed state is true or false. This applies to both domain-oriented processes and system security, where data-driven modeling, e.g. deployed in a security operations center, may reveal a high incidence of false alarms \cite{False_Alarms1}. We claim that a change in state is first perceived as such by a human expert and then, in a second stage, confirmed by associations and/or more in-depth contextual reasoning \cite{Change_Reasoning_1, Change_Reasoning_2, Situation_Aware_2} - as illustrated in Fig. \ref{fig:idea_intro}. The same is the case in algorithmics, where much effort has been devoted to identifying changes in patterns, such as structural change points \cite{maleki2016development, li2024automatic, matteson2014nonparametric, arlot2019kernel}, but hardly any work can be found on the statistical grounded authorization or negation of the occurrence of a given class of patterns (events) in the data. The difficulty further increases, and the efficiency and reliability of authorization decreases significantly for manually managed tasks when complex systems and processes and multi-sensor, multivariate streams of recorded signals are concerned - an example is the predictive maintenance service provided for manufacturing systems \cite{Practices_Predictive_Maintenance, CPD_Multivariate_2stage, Situation_Aware_1}.
 A recent industry review of more than 60 patents and research papers on anomaly detection in rotating machinery found that most false alarms in such systems result from fixed, manually set thresholds that fail to distinguish actual faults from normal operational variability, leading to alarm fatigue and an increased risk of overlooking real degradation \cite{PatSnap_False_Alarms}. As a result, user confidence in systems and processes implementing such mechanisms is fading. Mathematical statistics allows legitimizing the state of a random variable in terms of its significance \cite{myors2010statistical}. {The machine learning community addressing change point detection (CPD) typically evaluates performance by quantifying the error in identifying the time instant at which a data stream transitions between states. This error is commonly expressed as the number of samples or a time-normalized distance between the true and estimated change point indicators \cite{panek2022automated_change_point, panek_ASOC}, or alternatively by formulating the task as a classification problem and reporting standard classification metrics \cite{evaluation_cpd, Faber_Wasserstein}. Such evaluation protocols inherently assume the availability of ground-truth change point annotations, enabling the estimation of detection accuracy and false positive rates based on historical data. However, these approaches exhibit notable limitations in practical scenarios, for instance due to the inherent class imbalance, as change points constitute a minority class. In contrast, the approach proposed in this work does not aim at evaluating CPD methods under the assumption of known ground truth. Instead, it focuses on assessing the statistical significance of detected change points. Consequently, it provides a mechanism to determine whether a detected change point should indeed be regarded as a meaningful state transition or dismissed as statistically insignificant. This shifts the perspective from method evaluation toward increasing the certainty of results. Importantly, the proposed procedure can be applied online. Moreover, it enables the rejection of spurious detections, thereby increasing the reliability and interpretability of CPD outcomes rather than benchmarking the methods themselves. } Regarding the aforementioned, the challenge for the identified research gap is to design a significance test for multivariate and non-Gaussian random variables, an example of which in industrial applications may be a multi-sensor time series recording containing temporal impulsive distortions \cite{witulska2025identifying}. Solving this problem is relevant not only to algorithmic, but predominantly to the design of autonomous and trustworthy critical systems and processes \cite{Trustworthy_Autonomous_Critical_1, Trustworthy_Autonomous_Critical_2, Trustworthy_Autonomous_Critical_3}.

\subsection{Literature review}
{Change-point detection algorithms can be broadly categorized into online methods, enabling real-time detection and immediate system response to anomalies, and offline methods, which analyze pre-collected datasets to identify regime shifts retrospectively \cite{truong2020selective}. Both single and multiple change points may be targeted, with techniques further classified as parametric (assuming a specific distribution, typically Gaussian) or non-parametric. For univariate data, a wide range of CPD methods exists, often tailored to detecting specific types of changes such as shifts in mean, scale, trend, persistence, other distribution parameters or model coefficients \cite{chen2012monitoring, 2025robust, topal2016detecting, lee2016bayesian}. We also refer the readers to the position \cite{GRZESIEK2020108124}, where a methodology based on distribution-distances was proposed in the context of diagnostic data, and to \cite{maraj}, where the statistically-based CPD methodology was used as a preprocessing step for threshold line identification in the condition monitoring area.

{However, univariate analysis (also in the context of regime change or anomaly detection) may be insufficient to understand complex phenomena. For example, according to the assumptions of Medicine 4.0, solutions can be considered dedicated to Clinical Decision Support systems that assist specialists in decision-making in highly complex environments \cite{schnurr2018medicine, singh2023novel}. Multivariate CPD approaches are less developed but crucial, as they capture dependencies missed in single-variable analysis. Despite this, research often focuses on individual attributes, resulting in task-specific rather than general-purpose methodologies \cite{li2024automatic, matteson2014nonparametric, kalman2, ma2024change}. Existing methods built on diverse statistical frameworks (such as energy statistics, kernel-based tests, rank-based measures, likelihood estimators, copulas, or windowing techniques) are characterized by specific trade-offs in terms of efficiency, generality, and robustness to distributional assumptions \cite{matteson2014nonparametric, arlot2019kernel, witulska2025identifying, l1, l2}. {There exist also numerous non-statistical approaches for CPD, including machine learning methods \cite{li2022automatic, wang2021encoding}, graph-based methods \cite{aminikhanghahi2017survey}, and model-based methods \cite{janczura2023machine, chen2019inference}. The machine learning approaches to CPD have gained increasing attention, but they often depend on high-dimensional feature representations and can be challenging to apply effectively in real-time settings \cite{tian2025degradation}. On the other hand, advanced techniques leveraging statistical distance measures and state estimation frameworks, such as the Mahalanobis distance and Kalman filtering, have been widely explored \cite{kalman2, r4_5, r4_6, kalman1}. However, single-model approaches based on the Kalman filter may be insufficient to accurately capture the dynamics of complex systems \cite{imma}.} Although there are more general techniques for change point detection, their performance is often sensitive to the properties of the samples analyzed \cite{witulska2025identifying}. In particular, multidimensional CPD methods frequently degrade when the data depart from Gaussianity. For non-Gaussian heavy-tailed distributions, characterized by an increased probability of extreme values compared to the Gaussian case, the efficiency of many algorithms  decrease due to the strong influence of outliers on loss functions. Consequently, it becomes essential to validate detected change points using robust, outlier-resistant methods (so-called robust methods) to reduce false positives.}

 {In many machine learning tasks, such as classification, false positives correspond to incorrectly labeling an object as belonging to a given class. The false alarms reduce system reliability, trigger unnecessary responses, and increase analysis costs \cite{wang2015overview}. In the context of CPD, false positives occur when spurious change points are identified, leading to artificial partitioning of the signal and loss of coherence (physical consistency) in process interpretation. Strategies to reduce false positives primarily involve penalization and thresholding techniques (e.g., AIC/BIC criteria, minimal segment length) that prevent excessive signal splitting \cite{truong2020selective}. However, reliable detection without false alarms remains difficult because real-world signals are affected by random fluctuations, and distributional instabilities that can mimic true structural changes \cite{gold2018doubly}. Parametrization is particularly challenging, as it requires predefining values such as significance thresholds, window lengths, or penalization terms - parameters highly depends on the data’s distribution, noise level, and frequency of changes. Suboptimal choices may either inflate false positives or miss actual changes. Thus, striking a balance between sensitivity and specificity (and in case of online change point detection: with time) is a key challenge, with regard to the work on advanced statistical and adaptive approaches \cite{o2013revisiting, xu2025pushing}.

\subsection{Research gaps, contributions and novelty}
Based on the literature review and our own experience, the three research gaps have been identified: (i) {lack of statistical significance in segmentation algorithms.} The known CPD methods simply point out change without providing any statistical evidence, which is a major issue in real-world applications; (ii) {insufficient methodology for revealing uncertainties related to temporal parametric changes in complex data characteristics.} In multivariate real-world data, e.g.  that exhibit non-Gaussian heavy-tailed (impulsive) behavior, it is often uncertain which specific parameters of the distribution are changing. This makes it difficult to determine which data characteristics should be analyzed to assess the significance of a change point;
   (iii) {limited physics-awareness of the modeling used for multivariate industrial data with impulsive  distortions.} There is a gap in the ability to interpret and explain the identified and authorized change points which is significant in real-word scenarios. 

The novelty of this article is the introduction of \textit{Change point Hypothesis Assessment for Significance in Multivariate data} (CHASM), a novel non-parametric methodology for rigorously evaluating the significance of change points in complex time series, thus to authorize the occurrence of this pattern in industrial data, especially multivariate data exhibiting impulsive distortions. Unlike many existing CPD methods that lack a statistical foundation and can lead to false alarms, CHASM is designed to address such critical gaps. The method is highly scalable, adaptable to non-Gaussian distributions, and robust to outliers, thereby broadening its applicability. Furthermore, the novelty of CHASM extends beyond statistical testing, as it also facilitates a root-cause analysis of the detected changes, which is crucial for real-world applications where explainability is paramount for making informed business decisions and building trust in automated monitoring systems. {For instance, in industrial systems, the frequency of fault events (anomalies) is typically low, resulting in extremely limited fault samples. Consequently, many frequency-based methods becomes unreliable \cite{liu2026trustworthy, LIU2026_Trust_Review}. By evaluation detected change points and associates them with potential root causes, enabling interpretable system-level insights - CHASM facilitates effective knowledge extraction even in data-scarce conditions. 

Our contribution is also related to the presented case study which is based on a real-world dataset acquired on a reciprocating compressor in an oil and gas company. The company digitized its assets several years ago and it resulted in a large stream of data. Often, the amount of data exceeds the capacity of human experts, which are responsible for the compressor maintenance. The basic damage detection and identification tool are simple thresholds. Even so, setting threshold values is a non-trivial task. First, due to a very high number of monitored quantities and second, due to non-stationary operation of the machine, where changes in monitored parameters are much more often caused by a change of an operational state than by an incipient damage. Therefore, a method to detect early changes and assess its significance in a multidimensional process data is a very sought-after solution. To show the universality of the proposed methodology, we also demonstrate its performance for the vibration data from motorbike. The proposed approach can be widely adopted in industrial practice. 

The potential industrial applicability of CHASM is illustrated in Tab.~\ref{tab:chasm-industrial-risk}, which summarizes selected challenges encountered in real-world monitoring systems, their industrial implications, and their potential impact on system reliability and operational safety. The table also highlights how CHASM can address these challenges by supporting the assessment of detected changes, reducing false alarms, facilitating timely decision-making, and providing interpretable information that assists in identifying the underlying causes of system changes. By supporting more reliable change-point assessment, CHASM can contribute to improved risk evaluation, more effective maintenance decisions, and enhanced safety of industrial systems.

The rest of the article is organized as follows. In Section \ref{sec:methodology}, we introduce a novel general methodology (CHASM) for assessing the significance of the detected change points in multivariate data, together with the details related to the statistics used for this purpose. In Section \ref{sec:simulated_data}, we present the extended simulation studies for various scenarios based on multivariate Gaussian and Pareto distributions. In Section \ref{sec:real_data}, we outline how the designed approach can be used for change point authorization and discuss the significance of change points for multidimensional real-world industrial data and vibration data from motorbike. The last section concludes the article. In the Appendices, additional results are provided, including examples of simulated data, a comparison of the proposed approach with a baseline method reported in the literature for statistical assessment of the detected change point, an analysis of the impact of the window length on the results, and an evaluation of the robustness of the proposed approach based on simulations with additively noisy data.

\begin{table*}[!t]
\caption{Industrial consequences of change-point assessment errors and the role of CHASM in providing reliability of automated systems and operational safety.}
\label{tab:chasm-industrial-risk}
\centering
\footnotesize
\renewcommand{\arraystretch}{1.18}
\setlength{\tabcolsep}{3pt}
\begin{tabular}{@{}P{\ProblemColumnWidth}
                P{\IndustrialMeaningColumnWidth}
                P{\ConsequenceColumnWidth}
                P{\ChasmRelationColumnWidth}@{}}
\toprule
\textbf{Problem} &
\textbf{Industrial significance} &
\textbf{Consequence} &
\textbf{Relation to CHASM} \\
\midrule

False authorization &
An alarm is generated for a normal operating state. &
Unnecessary inspection; additional workload for maintenance personnel; time spent investigating false alarms. &
CHASM enables the definition of application-specific criteria for assessing the statistical significance of a detected change. By filtering statistically insignificant detections, it can reduce the number of false alarms. \\ \hline

False rejection &
An actual change in the system state is overlooked. &
Delayed response to a developing failure or degradation process. &
Change-point assessment involves a trade-off between false detections and missed changes. CHASM allows the decision threshold to be adapted to the application and provides criterion-specific evidence. The magnitude of the deviation and the number of criteria indicating a change can be considered when deciding which alarms require action or further investigation. \\ \hline

Delayed detection by algorithms based on complex functions or requiring training data &
The change is identified too late because the detection algorithm requires extensive computation, a long observation history, or a training phase. &
Reduced time available for a response before a failure. &
CHASM provides a multivariate, post-detection assessment that can be applied after a candidate change point has been identified. Because the assessment can use relatively simple statistics, computationally efficient detection algorithms operating on short windows may be combined with CHASM. The change can then be authorized using a richer, multi-criterion analysis. \\ \hline

Lack of a clear explanation of the cause of a change detected by a CPD algorithm &
The operator receives information that a change occurred, but not which variable, signal characteristic, or inter-component relationship changed. &
More difficult diagnosis; manual analysis of numerous signals; longer response time; and reduced confidence in the monitoring system. &
CHASM evaluates the change with respect to individual variables and criteria, including location, scale, distributional shape, inter-component dependence, and distributional distances. It identifies the variables and characteristics most strongly associated with the change, thereby supporting root-cause investigation. \\

\bottomrule
\end{tabular}
\end{table*}

\section{Methodology for statistical and explainable authorization}
\label{sec:methodology}

This section introduces a general methodology for evaluating the significance of change points detected by segmentation procedures, called CHASM\footnote{The code is published as GitHub repository by Zenodo platform (DOI: \url{https://doi.org/10.5281/zenodo.24499771} ).}. 
Initially, a detailed description of CHASM is provided. Next, we present few examples of statistics and divergences that can be used as the core of the proposed framework. Finally, we describe a baseline approach - a segmentation method that does not investigate the significance of the detected change points - for consideration in comparative analyses. 

\subsection{Mathematical model formulation}\label{math_problem}

Change-point detection refers to the procedure of partitioning $d$-dimensional independent random vectors $\mathbf{X}_1, \mathbf{X}_2, \ldots, \mathbf{X}_N$ $(N \in \mathbb{N})$ into portions (segments) that remain homogeneous within the segment \cite{Raphael1999360}. Homogeneity here means that all random vectors within a given segment have the same  distribution, i.e. the cumulative distribution function (CDF) of the sample within the segment is the same. Thus, the model is as follows:
  \begin{align}
\mathbf{X}_i \overset{d}{=} \begin{cases}
     \mathbf{X}^{(1)} \sim F_1, \quad &\text{for }  1 \leq i < n_1,\\
     \mathbf{X}^{(2)} \sim F_2, \quad &\text{for }  n_1 \leq i < n_2, \\
     ... \\
     \mathbf{X}^{(R)} \sim F_{R}, \quad &\text{for }  n_{R-1} \leq i < N,
    \end{cases}
    \label{eq:assumption_multiple_cp}
\end{align}
where $n_1, ..., n_{R-2}, n_{R-1}$ are the change points corresponding to splitting $S_N^R$ defined as follows:
\begin{align*}
    S_N^R = \{ (n_1, ..., n_{R-1}) \in \mathbb{N} ^ {R - 1} : 1 < n_1 < ... < n_{R-1} < N \}
\end{align*}
and $F_1, F_2, ..., F_{R-1}, F_{R}$ represent CDFs of $d$-dimensional random vectors corresponding to the following segments: $[1, n_1), [n_1, n_2), ..., [n_{R-2}, n_{R-1}), ... [n_{R-1}, N]$, respectively. This division (i.e. the identification of the change points) into segments is called segmentation. Let us mention that the model defined in (\ref{eq:assumption_multiple_cp}) is a general one discussed by many authors in the context of CPD \cite{witulska2025identifying}.

Importantly, segmentation algorithms rely on specific criteria (loss functions). Change points are most commonly detected based on such a criterion. In the case of algorithms designed for multiple CPD, additional stopping criteria or predefined thresholds are often introduced. However, these approaches face a major limitation - false positives, i.e., points identified because they minimize the loss function but which do not correspond to statistically significant changes in reality. 

The goal of this research is to introduce a method for assessing the statistical significance of detected change points, then discarding insignificant changes (false positives), and finally obtaining homogeneous data segments identified by significant change points. An additional functionality of the proposed approach is that it facilitates root-cause analysis of detected changes, making the segmentation results more explainable.

\subsection{Description of the procedure for testing the significance of changes}
\label{sec:proposed_approach}

The CHASM is a general, non-parametric method as it employs statistical testing using bootstrap techniques, thereby not requiring any prior knowledge of the distribution of the input data. Furthermore, the generality of CHASM is also due to the fact that it allows for customizable evaluation criteria, such as inter-component relationships or single components' properties (location, scale etc.). Consequently, a user with domain knowledge and a specific goal (the type of changes they wish to observe) can fine-tune the algorithm to support a precisely specified task.

Consider the multivariate data segmentation problem described in Section \ref{math_problem}. CHASM assesses the statistical significance of detected change points by combining segmentation with confidence interval testing. For each candidate change point, the data segments before and after the point are compared using a predefined set of statistics. A change is declared significant if the post-change statistic falls outside the confidence interval estimated from the pre-change segment. CHASM uses following input parameters: (i)
 window size ($\omega$) -- corresponds to the length of the samples taken for testing the significance of change points; (ii) number of resamples ($r$) -- number of bootstrap resamples used in algorithm; (iii) confidence level ($c$) -- is a proportion (e.g., 95\%) of confidence intervals, constructed from repeated random samples of the same population, that are expected to contain the true parameter value.

A key advantage of CHASM lies in its scalability, which stems from the structure of the input statistics. The method operates under the assumption that CHASM aggregates differences between statistics computed on marginal distributions and/or inter-component relationships, and on this basis determines whether a change in the multivariate distribution has occurred with respect to a given criterion(s). A criterion, or a set of criteria, may be tailored to the requirements of a specific application. In this work, we propose several criteria for characterizing changes in the data, including variations in mean, scale, dependence, skewness, and kurtosis, but this list could be expanded or modified depending on the user's needs. For the purposes of this study, we further introduce the concept of overall change, defined as the occurrence of a statistically significant difference in the underlying probability distributions (determined based on divergences measuring distances between distributions). Each of the proposed criteria can be formally expressed through statistical measures, as detailed in Section \ref{sec:statistics}. Here we call them as \textit{basic statistics}. What is essential, the mentioned \textit{basic statistics} are defined for univariate data. In order to apply them to the problem of assessing the significance of changes across multiple dimensions, it is necessary to introduce a transformation that aggregates the statistic values from all dimensions into a single scalar measure. 

For $\mathbf{X}_1, \mathbf{X}_2, ..., \mathbf{X}_N$ being  a  $d$-dimensional independent random vectors that follow the model given in Eq. (\ref{eq:assumption_multiple_cp}) we denote $\textbf{X}_{i:j} = \{\textbf{X}_i, \textbf{X}_{i+1}, ..., \textbf{X}_{j-1}, \textbf{X}_j \}$ for $i, j \in \mathbb{N}$ such that $i<j$. Let $X_{i:j}^k$ be sample of $k$-th components of $\textbf{X}_{i:j}$. CHASM assumes that this aggregated statistics can have one of the following two forms (Eq. \eqref{eq:stat1} and Eq. \eqref{eq:stat2}):
\begin{align}
    \mathcal{T}(T):= \max\limits_{k\in\{1,...,d\}}|T(X_{i:j}^k) - T(X_{s:t}^k) |, \label{eq:stat1}
\end{align}
\begin{equation}
    \mathcal{M}(M):= \max\limits_{m, k\in\{1,...,d\} \wedge
 m \neq k}|M(X_{i:j}^m, X_{i:j}^k) - M(X_{s:t}^m, X_{s:t}^k)|,
 \label{eq:stat2}
\end{equation}
where $j-i = t-s$ which ensures that the sub-samples used have the same lengths. Eq. \eqref{eq:stat1} represents a statistic that uses univariate \textit{basic statistics} $T(\cdot)$ present in the data (e.g., mean, scale, skewness, kurtosis, divergence measures), whereas Eq. \eqref{eq:stat2} uses a \textit{basic statistic} $M(\cdot, \cdot)$ characterizing the relationship between components (e.g., correlation coefficient). For more information about \textit{basic statistics} and their aggregation we used in this research see Section \ref{sec:statistics}. The selection of such statistics provides a broad insight into the characteristics of the data and enables the detection of regime shifts relevant to the real-world applications. For instance, a shift in the mean indicates a change in the baseline level of the process, a change in scale reflects a difference in process stability. Finally, a change in correlation between variables (here coordinates of the $d$-dimensional random variable) suggests altered coupling in the system.

 The idea of CHASM is visualized in the diagram (Fig. \ref{fig:idea_intro}) and the detailed description of the algorithm is as follows:

\begin{enumerate}
    \item Identify candidate change points in $\mathbf{X}_1, \mathbf{X}_2, \ldots, \mathbf{X}_N$ using the selected multivariate segmentation method.  

    \item Define a sequence of statistics $L_i \ (i = 1, 2, \ldots)$ (given by Eq. \eqref{eq:stat1} or Eq. \eqref{eq:stat2}), where each $L_i$ characterizes a specific type of distributional change to be detected and and finally recognized as significant.  

    \item For each detected change point $n^* \in \{n^*_1, n^*_2, \ldots\}$, evaluate its statistical significance with respect to each criterion $L_i$ according to the following procedure:  
    \begin{enumerate}
        \item Partition the segment $\mathbf{X}_{n^*-\omega:n^*}$ into two subsets (continuous split 50/50). Generate $r$ bootstrap resamples to obtain $r$ realizations of the statistic $L_i$. These bootstrap values form a reference distribution of $L_i$ under the null hypothesis of no distributional change.  
        \item Construct a one-sided confidence interval for the bootstrap distribution of $L_i$ at the confidence level $c$.  
        \item Compute the statistic $L_i$ for both $\mathbf{X}_{n^*-\omega:n^*}$ and $\mathbf{X}_{n^*+1:n^*+\omega+1}$, and assess whether the observed value falls within the constructed confidence interval. If the value lies outside the interval, then $n^*$ is marked as statistically significant with respect to criterion $L_i$.  
    \end{enumerate}

    \item A change point $n^*$ is regarded as a significant (true positive) if it is deemed significant for at least one statistic $L_i$. Otherwise, it is classified as not significant (false positive).  
\end{enumerate}

In Algorithm \ref{alg:significance} we present the pseudocode of the proposed procedure.

\begin{algorithm}[h!]
\caption{Change point Hypothesis Assessment for Significance in Multivariate data (CHASM)}
\label{alg:significance}
\KwIn{Multidimensional data $\bm{X}$, detected change points $\{n^*_1, n^*_2, \dots\}$, set of criteria $L$ (selected statistics), window length $\omega$, confidence level $c$, number of resamples $r$}
\KwOut{Set of statistically significant change points}

\BlankLine
\textbf{Step 1: Segmentation}\\
Apply a segmentation method to $\bm{X}$ to obtain candidate change points $\{n^*\}$.

\BlankLine
\textbf{Step 2: Window Selection}\\
Choose window length $\omega$, not greater than the length of the longest segment.

\BlankLine
\textbf{Step 3: Main of CHASM}\\
\ForEach{detected change point $n^*$}{
    Define intervals: \\
    \Indp
    $A = [n^* - \omega, \; n^*]$ \\
    $B = (n^*, \; n^* + \omega]$ \\
    \Indm
    
    \ForEach{criterion $L_i \in L$}{
        \textbf{(a)} Partition interval $A$ into two subsets (continuous division of the first segment in a 50/50 split) and compute one-sided confidence interval 
        for $L_i$, $c$, $r$. \\
        \textbf{(b)} Compute value $L_i(A, B)$ for data that correspond to intervals $A$ and $B$. \\
        \textbf{(c)} \eIf{$L_i(A, B)$ lies within the confidence interval}{
            Mark change point $n^*$ as \emph{not statistically significant} 
            with respect to criterion $L_i$.}
        {Mark change point $n^*$ as \emph{statistically significant} 
            with respect to criterion $L_i$.}
    }
}
\Return{Set of change points that are statistically significant for at least one $L_i$.}
\BlankLine
\end{algorithm}

\subsection{Description of the procedure for explainability assessment}
\label{sec:proposed_approach2}

For root-cause analysis, the framework of CHASM can be refined to the level of individual coordinates of $d-$dimensional vector, enabling the identification of whether a regime shift is directly associated with a specific variable. For this purpose, the algorithm described in Section~\ref{sec:proposed_approach} is applied independently to each component of the data (i.e., for each dimension $i \in \{1,...,d\}$). Consequently, the maximum values in formulas Eq. \eqref{eq:stat1} and Eq. \eqref{eq:stat2} are evaluated over a single dimension $d$, which naturally leads to their reduction. 
\subsection{Statistics and divergences used in the proposed methodology}
\label{sec:statistics}
In this section, we present the \textit{basic statistics} discussed in Section~\ref{sec:proposed_approach} and formally define their corresponding aggregate statistics, following the formulations in Eq. \eqref{eq:stat1} and Eq. \eqref{eq:stat2}. These aggregate measures are employed by the CHASM framework to evaluate the statistical significance of multivariate changes and are subsequently utilized in our analyses. 
We selected robust estimators to make CHASM independent of the data type, i.e., to ensure that for non-Gaussian data (particularly heavy-tailed distributions), the estimators are not biased due to the presence of outliers.

\subsubsection{L-moments}
L-moments are statistical measures used to summarize the shape of probability distributions. Unlike conventional moments (such as mean, variance, skewness, and kurtosis), L-moments are linear combinations of order statistics and offer increased robustness, especially in the presence of outliers or heavy-tailed distributions. L-moments are widely used in hydrology, climatology, and other fields that require robust parameter estimation for extreme value distributions, such as the Pareto distribution discussed in our simulation study. 

Let $X$ be a univariate random variable with CDF $F$. The $r$-th L-moment $\lambda_r$ is defined as follows \cite{lmoments}:

\begin{align}
\lambda_r = \frac{1}{r} \sum_{k=0}^{r-1} (-1)^k \binom{r-1}{k} \mathbb{E}[X_{r-k:r}],    
\end{align}
where $X_{i:r}$ denotes the $i$-th order statistic in a sample of size $r$ drawn from $F$. The first four L-moments are as follows:  $\theta_{L-LOC} = \lambda_1$ (L-location), $\theta_{L-SCALE} = \lambda_2$ (L-scale), $\theta_{L-SKEW} = \lambda_3 / \lambda_2$ (L-skewness),  $\theta_{L-KURT} = \lambda_4 / \lambda_2$ (L-kurtosis). 
In practice, to determine $\mathbb{E}[X_{r-k:r}]$ in the $\lambda_r$ formula ($r \in \{1,2,3,4\}$) for realizations of $X$, we use the sample mean defined as $\frac{1}{k}\sum_{i=r-k}^r x_i$.

\subsubsection{Robust estimator of scale}
Let $ X_1,...,X_N$ be a univariate random sample. The biweight midvariance is defined as follows \cite{lax1985robust}:
\begin{align}
    \theta_{BMID}  = N \ \frac{\sum_{|U_i| < 1} \
    (X_i - \hat{\mu}_r)^2 (1 - U_i^2)^4} {(\sum_{|U_i| < 1} \
    (1 - U_i^2) (1 - 5U_i^2))^2},
    \label{eq:BMID}
\end{align}
where $U_i = \frac{(X_i -\hat{\mu}_r )}{d \cdot MAD}$, $d$ is some constant and $MAD$ is {the median absolute deviation defined as} median of a sample $|X_1-\hat{\mu}_{r}|,|X_2-\hat{\mu}_{r}|,...,|X_N-\hat{\mu}_{r}|${, where $\hat{\mu}_{r}$ is a sample median of $X_1,...,X_N.$} 
As noted in \cite{lax1985robust}, the BMID estimator (and its variants) demonstrates superior performance compared to other robust scale estimators. 

\subsubsection{Jeffreys distance}
Jeffreys distance is a divergence metric between two probability distributions, introduced in \cite{jeffreys1946invariant}. For probability measures defined on a finite set of size $N$, represented as $P = (p_1, \ldots, p_N)$ and $Q = (q_1, \ldots, q_N)$, the Jeffreys distance is given by \cite{jeffrey}:
\begin{align}
D_J(P,Q) = \sum_{i=1}^{N} \left( \sqrt{p_i} - \sqrt{q_i} \right)^2.
\label{eq:dj}
\end{align}

\subsubsection{Bhattacharyya distance}
The Bhattacharyya distance is a symmetric measure of similarity between two probability distributions. For probability measures defined on a finite set of size $N$, represented as $P = (p_1, \ldots, p_N)$ and $Q = (q_1, \ldots, q_N)$, the Bhattacharyya distance is defined as \cite{bhattacharyya}:
\begin{eqnarray}
D_{\text{B}}(P, Q) = -\ln \left( \sum_{i=1}^N\sqrt{p_iq_i} \right).
\label{eq:db}
\end{eqnarray}
\noindent

\subsubsection{Spearman's rank correlation}
Spearman's rank correlation coefficient is a non-parametric measure of statistical dependence between two variables. It assesses how well the relationship between two variables can be described by a monotonic function. For given paired data $(Y_i, Z_i)$ for $i = 1, 2, \ldots, N$, each value is ranked and the coefficient is calculated as follows \cite{fieller1957tests}:
\begin{align}
    \theta_{CORR} = 1 - \frac{6 \sum \delta_i^2}{N(N^2 - 1)},
\end{align}
where \( \delta_i \) is the difference between the ranks of \( Y_i \) and \( Z_i \).

Spearman's correlation is robust to outliers and does not assume a linear relationship between variables, making it suitable for ordinal data or non-linear monotonic relationships.

\subsubsection{Aggregated statistics}
In this research, we employ divergences measuring distances between distributions, as well as robust estimators of mean, scale, skewness, and kurtosis, as the \textit{basic statistics} $T(\cdot)$ (see Eq. \eqref{eq:stat1}). Let us define the aggregate statistics used in the simulated study and real-data analyses: $\Theta_{L-LOC} = \mathcal{T}(\theta_{L-LOC})$, $\Theta_{L-SCALE} = \mathcal{T}(\theta_{L-SCALE})$, $\Theta_{L-SKEW} = \mathcal{T}(\theta_{L-SKEW})$, $\Theta_{L-KURT} = \mathcal{T}(\theta_{L-KURT})$, $\Theta_{BMID} = \mathcal{T}(\theta_{BMID})$, $D_{B} = \mathcal{T}(d_B)$, $D_{J} = \mathcal{T}(d_J)$. As the \textit{basic statistic} $M$ (see Eq. \eqref{eq:stat2}), we use Spearman’s correlation coefficient, which provides a robust estimator of correlation. Let us denote it $\Theta_{CORR} = \mathcal{M}(\theta_{CORR})$.

\subsection{Baseline algorithm for multivariate data segmentation}
\label{sec:baseline_method}

For comparative evaluation, we employ a widely recognized baseline method: Kernel Change Point Analysis (KCPA) \cite{arlot2019kernel}. KCPA is a kernel-based approach for detecting multiple change points in both univariate and multivariate data sequences. It leverages kernel measures to capture distributional shifts and applies a model selection strategy that optimizes a penalized least-squares criterion, thereby balancing model fit and segmentation complexity \cite{arlot2019kernel}. 

\section{Assessment of the proposed procedure for multivariate simulated data}
\label{sec:simulated_data}

{In this section, we evaluate the proposed CHASM procedure for multivariate data segmentation using computer simulations. Analyses described in Section \ref{sec:simulated_data} and Section \ref{sec:real_data} were performed on a machine with an AMD Ryzen 7 7730U with Radeon Graphics 2GHz, 32 GB RAM, running Windows 11 Pro. The code was written in Python 3.11, and executed in single-threaded mode.}

In our implementation, the discrete distributions in Eqs.~\eqref{eq:dj}–\eqref{eq:db} are obtained by quadrature on a uniform grid of 1000 points over the common interval.

{We present the experimental results obtained using simulated data containing either a single or multiple change points. In the experiments corresponding to the single change point scenario, we assume that the random vectors $\mathbf{X}_1, \ldots, \mathbf{X}_N \in \mathbb{R}^d$ follow the model defined by the segmentation assumption in Eq.~\eqref{eq:assumption_multiple_cp}, with $R = 2$:}
\begin{align}
\mathbf{X}_i \overset{d}{=} \begin{cases}
     \mathbf{X}^{(1)} \sim F_1, \quad \text{for } i < n^*,\\
     \mathbf{X}^{(2)} \sim F_2 \quad \text{for } i \geq n^*,
    \end{cases}
    \label{eg:assumption_one_cp}
\end{align}
where $n^*$ is a change point, $F_1, F_2$ represent distributions of $d$-dimensional random vectors for the first segment and for the second segment, respectively. 
We evaluate the effectiveness of CHASM through extensive Monte Carlo simulations on multivariate Gaussian and Pareto distributions, with performance assessed in terms of statistical power, Type~I error, and computational efficiency. Such multivariate distributions are selected to cover both classical (Gaussian) and not-classical (heavy-tailed) distributions. In addition, they are widely applied in different areas of interest. {Representative trajectories from both distributions can be found in \ref{app:simulated_data}.}  

For two class of multivariate distributions, we also perform a comparative evaluation between the baseline method (i.e., pure KCPA algorithm described in Section \ref{sec:baseline_method}) and the CHASM used on KCPA detected change points generated from model (\ref{eq:assumption_multiple_cp}) for $R>2$. 

Moreover, \ref{app:likelihood} compares the accuracy and power of CHASM with those of the parametric likelihood-ratio test under scenarios analogous to those considered in this section, for both the multivariate Gaussian and multivariate Pareto distributions. The results presented in this chapter and in Appendix B correspond to procedures employing a window length of $\omega = 200$.

\ref{sec:window_size} presents the results obtained for the multivariate Gaussian distribution using different window lengths. In addition, this section provides practical guidelines for selecting an appropriate window length based on the characteristics of the data and the statistical tests employed within the proposed methodology.

As the Pareto distribution is characterized by heavy tails, the high effectiveness of CHASM in this setting provides evidence of the robustness of the proposed methodology to heavy-tailed data. However, to further demonstrate its robustness to other types of distributional changes, such as changes in the dependence structure, \ref{app:cont_gauss} evaluates CHASM on contaminated multivariate Gaussian data. This analysis also illustrates the detection limits of the considered statistical tests, particularly in terms of the magnitude of changes that can be detected in the presence of noise.

\subsection{Multivariate Gaussian distribution}
Gaussian (or normal) distribution plays a crucial role in probability theory and statistics due to their analytical properties and prevalence in natural and social phenomena. It is commonly used in hypothesis testing due to the Central Limit Theorem \cite{clt}. A univariate random variable $X$ has Gaussian distribution, when its probability density function (PDF) is as follows: 
\begin{align}
{\displaystyle f(x) = \frac{\exp\left( -\frac{(x - \mu)^2}{2\sigma^2} \right)}{\sqrt{2\pi\sigma^2}}, \quad x \in \mathbb{R},}
\label{eq:gauss_onedim_pdf}
\end{align}
\noindent
where 
$\mu > 0$ represents mean (location), and $\sigma$ is a scale.

The multivariate Gaussian distribution generalizes the univariate case to $d$ dimensions. A random vector $\mathbf{X} \in \mathbb{R}^d$ is said to follow a multivariate Gaussian distribution if its PDF is as follows: 
\begin{align}
{\displaystyle f(\mathbf{x}) = \frac{\exp\left( -\frac{1}{2} (\mathbf{x} - \bm{\mu})^\top \bm{\Sigma}^{-1} (\mathbf{x} - \bm{\mu}) \right)}{(2\pi)^{d/2} |\bm{\Sigma}|^{1/2}}, \quad \mathbf{x} \in \mathbb{R}^d,}
\label{eq:gauss_multidim_pdf}
\end{align}
\noindent
where 
$\bm{\mu}$ represents $d$-dimensional mean vector, and $\bm{\Sigma} = \{\sigma_i\sigma_j\rho_{ij}\}_{i,j=1}^d$ is $d \times d$ covariance matrix. In this article, we take the notation $\mathbf{X} \sim G_{\bm{\mu}, \bm{\Sigma}, d}$. Let us note, that $\sigma_i^2$ is the variance of the $i-$th coordinate of the vector $\mathbf{X}$ while $\rho_{ij}$ is the correlation coefficient between $i$-th and $j-$th coordinate of $\mathbf{X}$.

In the simulations presented  we consider $d$-dimensional vectors $\mathbf{X}_1,...,\mathbf{X}_N \in \mathbb{R}^d$ that correspond to the model \eqref{eg:assumption_one_cp}, where:
\begin{align}
\mathbf{X}_i \overset{d}{=} \begin{cases}
     \mathbf{X}^{(1)} \sim G_{\bm{\mu}_1, \bm{\Sigma}_1, d}, \quad \text{for } i < n^*,\\
     \mathbf{X}^{(2)} \sim G_{\bm{\mu}_2, \bm{\Sigma}_2, d} \quad \text{for } i \geq n^*.
    \end{cases}
    \label{eg:gaussian_one_cp}
\end{align}

\noindent
For the first segment (i.e., for $i < n^*$), we set 
$\bm{\mu}_1 = (1, \ldots, 1)$ and $\bm{\Sigma}_1 = I_d$, where  $I_d$ is identity $d \times d$ matrix while for the second segment (i.e., for $i \geq n^*$), we assume that the variances of the coordinates are homogeneous, that is, 
$\sigma_i^2 = \sigma^2$ for all $i \in \{1, \ldots, d\}$. 
The correlation structure is specified such that 
$\rho_{i,j} = 0$ whenever $\{i,j\} \notin \{\{1,2\}, \{1,3\}, \{2,3\}\}$ and $i \neq j$. 
The remaining values of $\bm{\mu}_2$ and $\bm{\Sigma}_2$, vary depending on the specific scenarios considered in the simulations. 
A detailed description of these scenarios is provided in Tab.~\ref{tab:experiments_gauss}.

\begin{table}[h!]
\centering
\caption{Parameters of $\mathbf{X}^{(2)}$ described in model \eqref{eg:gaussian_one_cp}.}
\label{tab:experiments_gauss}
\begin{tabular}{lllllll}
\hline
\textbf{id} & \textbf{factor} & \textbf{$\bm{\mu}_2$} & \textbf{$\sigma^2$} & \textbf{$\rho_{12}$} & \textbf{$\rho_{13}*$} & \textbf{$\rho_{23}*$} \\ \hline
1                      & $\bm{\mu}$             & $\{2,1,…,1\}$         & 1                   & 0                    & 0                    & 0                    \\
2                      & $\bm{\mu}$             & $\{2,2,…,2\}$         & 1                   & 0                    & 0                    & 0                    \\
3                      & $\sigma$               & $\{1,1,…,1\}$         & 2                   & 0                    & 0                    & 0                    \\
4                      & $\rho$                 & $\{1,1,…,1\}$         & 1                   & 0.9                  & 0.3                  & 0.2                  \\
5                      & $\rho$                 & $\{1,1,…,1\}$         & 1                   & 0.5                  & 0.3                  & 0.2                  \\ \hline
\end{tabular}
\caption*{*The values of $\rho_{12}$ and $\rho_{23}$ are defined only for $d>2$. }
\end{table}

For each experiment described in Tab. \ref{tab:experiments_gauss}, we conduct 100 Monte Carlo simulations of the random vectors described by Eq. \eqref{eg:gaussian_one_cp} for which $d \in \{2, 3, 5, 10, 25\},$ $n^*=200$, $N=400$. The significance level used in statistical tests is set to 5\%. In Tab. \ref{tab:power_gauss} we can find the values of power of the statistical tests based on selected statistics 
described in the Section \ref{sec:methodology}. The power of the test is the probability that the test correctly rejects the null hypothesis (change point is not significant based on selected criterion) when the alternative hypothesis (change point is significant based on selected criterion) is true. It is calculated as follows:
\begin{align}
    \hat{\mathcal{P}} = \frac{\#m \in \{1,...,M\}: \text{change is significant}}{M},
\end{align}
where $M$ is a number of the Monte Carlo trials.

\begin{table*}[t!]
\caption{Power analysis of CHASM for multivariate Gaussian data.}
\label{tab:power_gauss}
\centering
\begin{tabular}{llcclllllllll}
\hline
\textbf{no} & \textbf{d} & \textbf{change factor} & \textbf{id} & \textbf{$\Theta_{L-LOC}$}               & \textbf{$\Theta_{L-SCALE}$}                  & \textbf{$\Theta_{BMID}$}            & \textbf{$\Theta_{L-SKEW}$}  & \textbf{$\Theta_{L-KURT}$}  & \textbf{$\Theta_{CORR}$}           & \textbf{$D_B$}  & \textbf{$D_J$} & $\mathcal{R}$       \\ \hline
1  & 2  &                     &             & \cellcolor[HTML]{EFEFEF}0.64 & 0                            & 0                            & 0             & 0             & 0                            & \cellcolor[HTML]{EFEFEF}0.97 & \cellcolor[HTML]{EFEFEF}0.99 & 0.99 \\
2  & 3  &                  &                & \cellcolor[HTML]{EFEFEF}0.53 & 0                            & 0                            & 0             & 0             & 0                            & \cellcolor[HTML]{EFEFEF}0.93 & \cellcolor[HTML]{EFEFEF}0.98 & 0.98 \\
3  & 5  &               &                   & \cellcolor[HTML]{EFEFEF}0.51 & 0                            & 0                            & 0             & 0             & 0                            & \cellcolor[HTML]{EFEFEF}0.82 & \cellcolor[HTML]{EFEFEF}0.96 & 0.96 \\
4  & 10 &                   &               & \cellcolor[HTML]{EFEFEF}0.30  & 0                            & 0                            & 0             & 0             & 0                            & \cellcolor[HTML]{EFEFEF}0.71 & \cellcolor[HTML]{EFEFEF}0.91 & 0.91 \\
5  & 25 & \multirow{-5}{*}{$\bm{\mu}$} & \multirow{-5}{*}{1}     & \cellcolor[HTML]{EFEFEF}0.18 & 0                            & 0                            & 0             & 0             & 0                            & \cellcolor[HTML]{EFEFEF}0.60  & \cellcolor[HTML]{EFEFEF}0.84 & 0.84 \\ \hline
6  & 2  &               &                   & \cellcolor[HTML]{EFEFEF}0.82 & 0                            & 0                            & 0             & 0             & 0                            & \cellcolor[HTML]{EFEFEF}1    & \cellcolor[HTML]{EFEFEF}1  & 1  \\
7  & 3  &                    &              & \cellcolor[HTML]{EFEFEF}0.85 & 0                            & 0                            & 0             & 0             & 0                            & \cellcolor[HTML]{EFEFEF}1    & \cellcolor[HTML]{EFEFEF}1  & 1  \\
8  & 5  &                  &                & \cellcolor[HTML]{EFEFEF}0.91 & 0                            & 0                            & 0             & 0             & 0                            & \cellcolor[HTML]{EFEFEF}1    & \cellcolor[HTML]{EFEFEF}1  & 1  \\
9  & 10 &                  &                & \cellcolor[HTML]{EFEFEF}0.92 & 0                            & 0                            & 0             & 0             & 0                            & \cellcolor[HTML]{EFEFEF}1    & \cellcolor[HTML]{EFEFEF}1  & 1  \\
10 & 25 & \multirow{-5}{*}{$\bm{\mu}$} & \multirow{-5}{*}{2}     & \cellcolor[HTML]{EFEFEF}0.90  & 0                            & 0                            & 0             & 0             & 0                            & \cellcolor[HTML]{EFEFEF}1    & \cellcolor[HTML]{EFEFEF}1 & 1   \\ \hline
11 & 2  &                   &               & 0.01                         & \cellcolor[HTML]{EFEFEF}0.86 & \cellcolor[HTML]{EFEFEF}0.95 & 0             & 0             & 0                            & 0.03                         & 0.14                  & 0.95       \\
12 & 3  &                 &                 & 0                            & \cellcolor[HTML]{EFEFEF}0.82 & \cellcolor[HTML]{EFEFEF}0.91 & 0             & 0             & 0                            & 0                            & 0.14    & 0.91                      \\
13 & 5  &              &                    & 0                            & \cellcolor[HTML]{EFEFEF}0.86 & \cellcolor[HTML]{EFEFEF}0.93 & 0             & 0             & 0                            & 0                            & 0.08   & 0.93                       \\
14 & 10 &              &                    & 0                            & \cellcolor[HTML]{EFEFEF}0.81 & \cellcolor[HTML]{EFEFEF}0.93 & 0             & 0             & 0                            & 0                            & 0.06    & 0.93                      \\
15 & 25 & \multirow{-5}{*}{$\sigma$} & \multirow{-5}{*}{3}     & 0                            & \cellcolor[HTML]{EFEFEF}0.93 & \cellcolor[HTML]{EFEFEF}0.93 & 0             & 0             & 0                            & 0                            & 0          & 0.93                   \\ \hline
16 & 2  &                 &                 & 0                            & 0                            & 0                            & 0             & 0             & \cellcolor[HTML]{EFEFEF}1    & 0                            & 0                   & 1          \\
17 & 3  &                      &            & 0                            & 0                            & 0                            & 0             & 0             & \cellcolor[HTML]{EFEFEF}1    & 0                            & 0                     & 1        \\
18 & 5  &                   &               & 0                            & 0                            & 0                            & 0             & 0             & \cellcolor[HTML]{EFEFEF}1    & 0                            & 0                   & 1          \\
19 & 10 &                    &              & 0                            & 0                            & 0                            & 0             & 0             & \cellcolor[HTML]{EFEFEF}1    & 0                            & 0                     & 1        \\
20 & 25 & \multirow{-5}{*}{$\rho$} & \multirow{-5}{*}{4}     & 0                            & 0                            & 0                            & 0             & 0             & \cellcolor[HTML]{EFEFEF}1    & 0                            & 0                  & 1           \\ \hline
21 & 2  &                &                  & 0                            & 0                            & 0                            & 0             & 0             & \cellcolor[HTML]{EFEFEF}1    & 0                            & 0                       & 1      \\
22 & 3  &                 &                 & 0                            & 0                            & 0                            & 0             & 0             & \cellcolor[HTML]{EFEFEF}0.96 & 0                            & 0                       & 0.96     \\
23 & 5  &                 &                 & 0                            & 0                            & 0                            & 0             & 0             & \cellcolor[HTML]{EFEFEF}0.97 & 0                            & 0                      & 0.97      \\
24 & 10 &                 &                 & 0                            & 0                            & 0                            & 0             & 0             & \cellcolor[HTML]{EFEFEF}0.78 & 0                            & 0                   & 0.78         \\
25 & 25 & \multirow{-5}{*}{$\rho$} & \multirow{-5}{*}{5}     & 0                            & 0                            & 0                            & 0             & 0             & \cellcolor[HTML]{EFEFEF}0.43 & 0                            & 0            & 0.43                \\ \hline
\end{tabular}
\end{table*}

\begin{table*}[t!]
\caption{Type~I error for CHASM for multivariate Gaussian data.}
\label{tab:first_error_gauss}
\centering
\begin{tabular}{llllllllll}
\hline
\textbf{no} & \textbf{d} &  \textbf{$\Theta_{L-LOC}$}  & \textbf{$\Theta_{L-SCALE}$}  & \textbf{$\Theta_{BMID}$} & \textbf{$\Theta_{L-SKEW}$}  & \textbf{$\Theta_{L-KURT}$}  & \textbf{$\Theta_{CORR}$} & \textbf{$D_B$} & \textbf{$D_J$} \\ \hline
1           & 2                      & 0               & 0                & 0                     & 0                   & 0                   & 0                      & 0                           & 0                     \\
2           & 3                        & 0               & 0                & 0                     & 0                   & 0                   & 0                      & 0                           & 0                     \\
3           & 5                        & 0               & 0                & 0                     & 0                   & 0                   & 0                      & 0                           & 0                     \\
4           & 10                       & 0               & 0                & 0                     & 0                   & 0                   & 0                      & 0                           & 0                     \\
5           & 25                     & 0               & 0                & 0                     & 0                   & 0                   & 0                      & 0                           & 0                     \\ \hline
\end{tabular}
\end{table*}

For each simulation, we treated the first 200 observations as the first segment, and the last 200 observations as the second segment taken for two-sample tests used in CHASM. As presented in Tab.~\ref{tab:power_gauss}, for both experiment~1 and experiment~2 (see first two rows in Tab. \ref{tab:experiments_gauss}), the test based on $\Theta_{L-LOC}$ (see notations from Section \ref{sec:methodology}) detects significant changes in $\bm{\mu}$ in the majority of cases. This holds even when the mean shift occurs in only a single component out of $d$ possible dimensions, although in such cases the test power decreases noticeably as the dimensionality $d$ increases. Furthermore, tests based on both $D_J$ and $D_B$ also indicate the presence of a significant change. This outcome is consistent with expectations, as these criteria are designed to be highly sensitive to alterations in the marginal distributions of the analyzed data.

In the case of a scale change (experiment~3), the tests based on $\Theta_{L-SCALE}$ and $\Theta_{BMID}$ successfully detect significant differences in scale, consistent with the true changes introduced in the experiment. Specifically, the test based on $\Theta_{L-SCALE}$ achieves a statistical power in the range of 0.81--0.93, depending on the dimensionality $d$, while the test based on $\Theta_{BMID}$ attains a power of 0.91--0.95 across the same range of dimensions. For experiments involving changes in the correlation structure of the data (experiments~4 and~5), the tests based on $D_J$ and $D_B$, which are computed from univariate distributions, do not indicate any significant change. This result is consistent with expectations, as the modifications pertain to the dependence structure among variables rather than to the characteristics of individual components. As can be seen, in this case the test based on $\Theta_{CORR}$ has mostly a high power indicating significant changes in the correlation.

Suppose that we want to use CHASM to assess the significance of detected change point when change type is uknown (i.e., we assume that change point can be any type e.g., mean, scale, correlation, or other parameters). 
We define the change point as significant if at least one of the applied criteria ($\Theta_{L-LOC}, \Theta_{L-SCALE}, \Theta_{BMID}, \Theta_{L-SKEW}, \Theta_{L-KURT}$, $\Theta_{CORR}, D_J$ and $D_B$) detects a statistically significant change. {We define $\mathcal{R}$ as \textit{trustworthy factor} that measures the accuracy of our procedure in the simulated scenarios. This factor -- that is located in last column  of Tab.~\ref{tab:power_gauss} -- reports the proportion of simulations in which at least one of the considered criteria identified a change.} As observed, for the majority of experiments and across different dimensionalities $d$, the accuracy of detecting a significant change exceeds 90\%.

\begin{figure}[h!]
    \centering
    \includegraphics[width=0.9\linewidth]{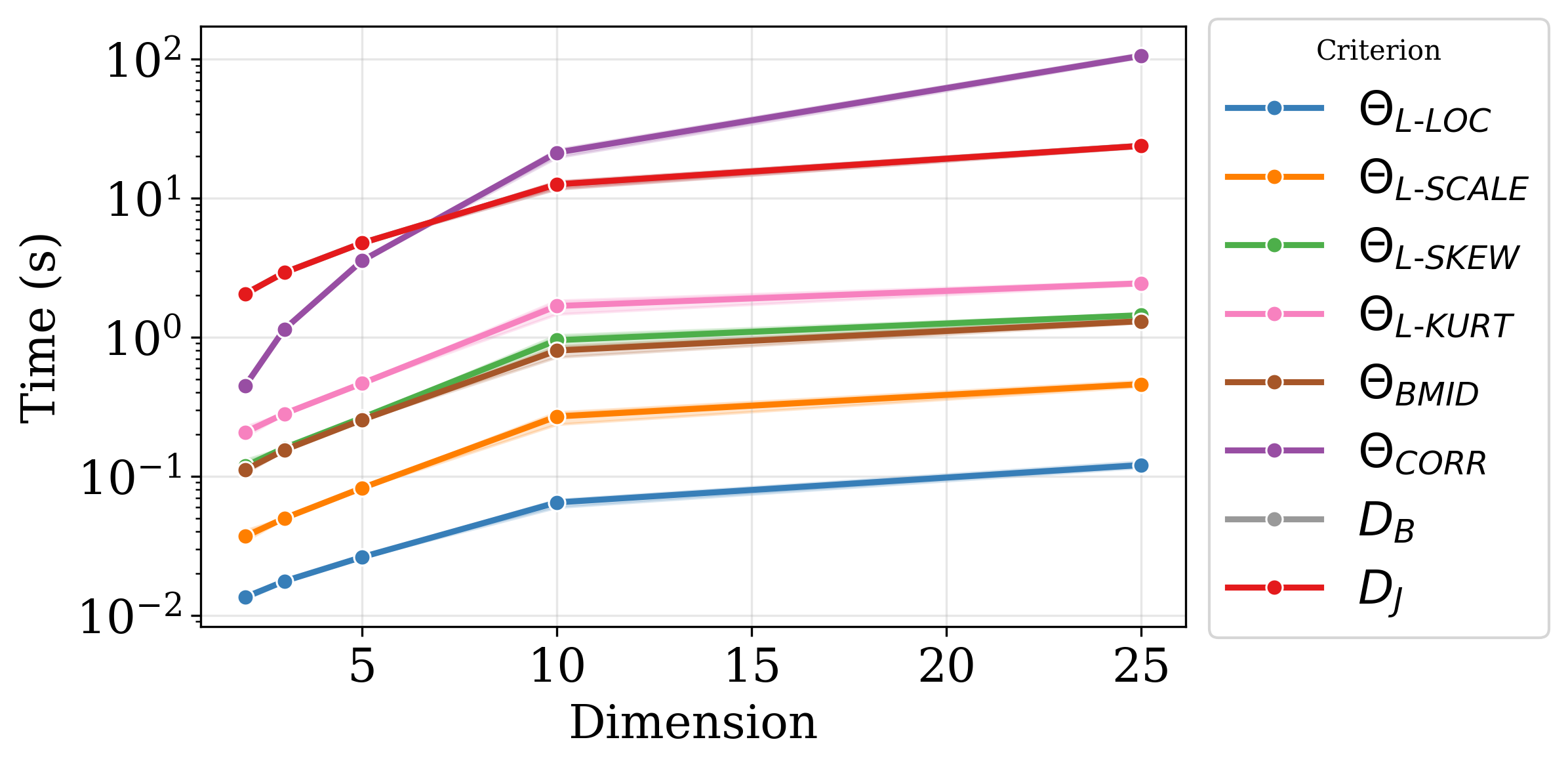}
    \caption{Runtime computational complexity for selected statistics used by CHASM for multivariate Gaussian distributed data.}
    \label{fig:gauss_times}
\end{figure}

Finally, we evaluate the Type~I error associated with the statistical tests employed by CHASM. {We recall, in statistics the Type~I error occurs  when rejecting the null hypothesis $H_0$ even though it is actually true. In our case, the null hypothesis is that selected change point is not significant based on selected criterion(s). Here, our aim is to check how many times the algorithm indicates a significant change between two samples while they are of the same distribution.} For this purpose, we simulate data of length $N = 400$ from the distribution $G_{\bm{\mu}_1, \bm{\Sigma}_1, d}$, that is, without any structural changes. Across 100 Monte Carlo replications, the CHASM procedure was applied by treating the first half of the data as the first segment and the second half as the second segment.

The corresponding results are reported in Tab.~\ref{tab:first_error_gauss}. As observed, for every considered dimensionality $d$, the outcome is correct, with no statistically significant change detected. Consequently, irrespective of the combination of criteria integrated within CHASM, the procedure does not produce any false positives in these simulations.

We also verify the computational complexity at runtime for selected tests performed within the CHASM procedure. The results are presented in Fig. \ref{fig:gauss_times}.

\subsection{Multivariate Pareto distribution}
\label{sec:pareto}
A random vector $\mathbf{X} \in \mathbb{R}^d$ is said to follow Pareto Type II distribution (called later Pareto distribution) if its PDF is as (see e.g. \cite{arendarczyk2018}):
\begin{align}
\displaystyle f(\mathbf{x}) = \frac{(\alpha\beta)^d \Gamma(1/\alpha + d)}{\Gamma(1/\alpha)} \left(1 + \alpha\beta \sum_{i=1}^{d} x_i \right)^{-1/\alpha - d},
\label{eq:pareto_pdf}
\end{align}
\noindent
where $x_i>0$ ($i=1,2,\ldots,d$),
$\alpha \geq 0$ represents a tail parameter, and $\beta > 0$ is a scale parameter. We recall, when $\alpha<0.5$, the variance of the coordinates of the multivariate Pareto distribution is finite while it is infinite otherwise. In this article, we take the notation $\mathbf{X} \sim MP_{\alpha, \beta, d}$. Let us recall that the multivariate Pareto distributed random vector $\mathbf{X}$ is represented as follows \cite{hogg2013introduction}:

\begin{equation}
\mathbf{X} \overset{d}{=} \frac{1}{Z} (E_1, E_2, \ldots, E_d),
\end{equation}
where $\mathbf{E} = (E_1, E_2, \ldots, E_d)$ is a vector of independent and identically distributed (i.i.d.) exponential random variables, with the following PDF:
\begin{equation}
f(x) = \beta e^{-\beta x}, \quad x > 0
\end{equation}
and $Z$ is a positive random scale factor, that is a Gamma random variable, independent of $\mathbf{E}$, with the following PDF:

\begin{equation}
f(x) = \frac{(1/\alpha)^{1/\alpha} x^{1/\alpha - 1} e^{-x/\alpha}}{\Gamma(1/\alpha)}, \quad x > 0.
\end{equation}

In the simulations presented in the following part we consider $d$-dimensional vectors $\mathbf{X}_1,...,\mathbf{X}_N \in \mathbb{R}^d$ that correspond to the model \eqref{eg:assumption_one_cp}, where:
\begin{align}
\mathbf{X}_i \overset{d}{=} \begin{cases}
     \mathbf{X}^{(1)} \sim MP_{\alpha_1, \beta_1, d}, \quad \text{for } i < n^*,\\
     \mathbf{X}^{(2)} \sim MP_{\alpha_2, \beta_2, d} \quad \text{for } i \geq n^*.
    \end{cases}
    \label{eg:pareto_one_cp}
\end{align}

\noindent
Detailed information on the parameters used in the experiments is provided in Tab.~\ref{tab:experiments_pareto}.
For each experiment we conduct 100 Monte Carlo simulations of the random vectors described by Eq. \eqref{eg:pareto_one_cp} for $d \in \{2, 3, 5, 10, 25\},$ $n^*=200$, $N=400$. The significance level used in tests is set to 5\%. In the Tab. \ref{tab:power_pareto} we can find the values of power of the statistical tests for significance of the change point based on selected statistics described in the Section \ref{sec:methodology}.

\begin{table}[t!]
\caption{Parameters of model \eqref{eg:pareto_one_cp} used in simulations.}
\label{tab:experiments_pareto}
\centering
\begin{tabular}{llllll}
\hline
\textbf{id} & \textbf{change factor} & \textbf{$\alpha_1$} & \textbf{$\alpha_2$} & \textbf{$\beta_1$} & \textbf{$\beta_2$} \\ \hline
6                      & $\alpha$               & 0.4                 & 0.8                 & 1                  & 1                  \\
7                      & $\alpha$               & 0.2                 & 0.4                 & 1                  & 1                  \\
8                      & $\beta$                & 0.4                 & 0.4                 & 1                  & 10                 \\
9                      & $\beta$                & 0.4                 & 0.4                 & 0.2                & 10                 \\ \hline
\end{tabular}
\end{table}

\begin{table*}[t!]
\caption{Power analysis of CHASM for multivariate Pareto data.}
\label{tab:power_pareto}
\centering
\begin{tabular}{llcclllllllll}
\hline
\textbf{no} & \textbf{d} & \textbf{change factor} &  \textbf{id} &\textbf{$\Theta_{L-LOC}$}  & \textbf{$\Theta_{L-SCALE}$}              & \textbf{$\Theta_{BMID}$}     & \textbf{$\Theta_{L-SKEW}$}  & \textbf{$\Theta_{L-KURT}$}  & \textbf{$\Theta_{CORR}$} & \textbf{$D_B$}  & \textbf{$D_J$}  & $\mathcal{R}$      \\ \hline
1  & 2  &                 &                   & 0.04             & \cellcolor[HTML]{EFEFEF}1    & \cellcolor[HTML]{EFEFEF}1 & 0.32              & 0.34              & 0.04            & \cellcolor[HTML]{EFEFEF}0.98 & \cellcolor[HTML]{EFEFEF}1  & 1   \\
3  & 3  &                    &                & 0                & \cellcolor[HTML]{EFEFEF}0.98 & \cellcolor[HTML]{EFEFEF}1 & 0.24              & 0.32              & 0               & \cellcolor[HTML]{EFEFEF}0.94 & \cellcolor[HTML]{EFEFEF}0.94 & 1 \\
5  & 5  &                 &                   & 0.02             & \cellcolor[HTML]{EFEFEF}0.98 & \cellcolor[HTML]{EFEFEF}1 & 0.16              & 0.28              & 0               & \cellcolor[HTML]{EFEFEF}0.98 & \cellcolor[HTML]{EFEFEF}0.98 & 1 \\
7  & 10 &                &                    & 0                & \cellcolor[HTML]{EFEFEF}1    & \cellcolor[HTML]{EFEFEF}1 & 0.3               & 0.34              & 0               & \cellcolor[HTML]{EFEFEF}1    & \cellcolor[HTML]{EFEFEF}1 & 1    \\
9  & 25 & \multirow{-5}{*}{$\alpha$} & \multirow{-5}{*}{1}     & 0                & \cellcolor[HTML]{EFEFEF}1    & \cellcolor[HTML]{EFEFEF}1 & 0.12              & 0.22              & 0               & \cellcolor[HTML]{EFEFEF}0.96 & \cellcolor[HTML]{EFEFEF}0.94 & 1 \\ \hline
2  & 2  &                 &                   & 0                & \cellcolor[HTML]{EFEFEF}1    & \cellcolor[HTML]{EFEFEF}1 & 0.04              & 0.06              & 0               & \cellcolor[HTML]{EFEFEF}1    & \cellcolor[HTML]{EFEFEF}1 & 1    \\
4  & 3  &                      &              & 0.02             & \cellcolor[HTML]{EFEFEF}1    & \cellcolor[HTML]{EFEFEF}1 & 0.06              & 0.1               & 0               & \cellcolor[HTML]{EFEFEF}1    & \cellcolor[HTML]{EFEFEF}1  & 1  \\
6  & 5  &                 &                   & 0                & \cellcolor[HTML]{EFEFEF}1    & \cellcolor[HTML]{EFEFEF}1 & 0.06              & 0.06              & 0               & \cellcolor[HTML]{EFEFEF}1    & \cellcolor[HTML]{EFEFEF}1  & 1  \\
8  & 10 &                 &                   & 0                & \cellcolor[HTML]{EFEFEF}1    & \cellcolor[HTML]{EFEFEF}1 & 0.06              & 0.1               & 0               & \cellcolor[HTML]{EFEFEF}1    & \cellcolor[HTML]{EFEFEF}1  & 1  \\
10 & 25 & \multirow{-5}{*}{$\alpha$} & \multirow{-5}{*}{2} & 0                & \cellcolor[HTML]{EFEFEF}1    & \cellcolor[HTML]{EFEFEF}1 & 0.02              & 0.04              & 0               & \cellcolor[HTML]{EFEFEF}1    & \cellcolor[HTML]{EFEFEF}1 & 1   \\ \hline
11 & 2  &                  &                  & 0                & 0.22                         & 0                         & 0.02              & 0.02              & 0               & \cellcolor[HTML]{EFEFEF}1    & \cellcolor[HTML]{EFEFEF}1  & 1  \\
13 & 3  &                  &                  & 0                & 0.18                         & 0                         & 0                 & 0                 & 0               & \cellcolor[HTML]{EFEFEF}1    & \cellcolor[HTML]{EFEFEF}1  & 1  \\
15 & 5  &                 &                   & 0                & 0.04                         & 0                         & 0                 & 0                 & 0               & \cellcolor[HTML]{EFEFEF}0.98 & \cellcolor[HTML]{EFEFEF}1  & 1  \\
17 & 10 &                 &                   & 0                & 0                            & 0                         & 0                 & 0                 & 0               & \cellcolor[HTML]{EFEFEF}0.98 & \cellcolor[HTML]{EFEFEF}1  & 1  \\
19 & 25 & \multirow{-5}{*}{$\beta$} & \multirow{-5}{*}{3} & 0                & 0                            & 0                         & 0                 & 0.04              & 0               & \cellcolor[HTML]{EFEFEF}1    & \cellcolor[HTML]{EFEFEF}1  & 1  \\ \hline
12 & 2  &                  &                  & 0                & 0.26                         & 0                         & 0                 & 0                 & 0               & \cellcolor[HTML]{EFEFEF}1    & \cellcolor[HTML]{EFEFEF}1  & 1  \\
14 & 3  &                 &                   & 0                & 0.1                          & 0                         & 0.02              & 0.02              & 0               & \cellcolor[HTML]{EFEFEF}1    & \cellcolor[HTML]{EFEFEF}1  & 1  \\
16 & 5  &                  &                  & 0                & 0.12                         & 0                         & 0                 & 0                 & 0               & \cellcolor[HTML]{EFEFEF}1    & \cellcolor[HTML]{EFEFEF}1  & 1  \\
18 & 10 &                  &                  & 0                & 0.06                         & 0                         & 0                 & 0                 & 0               & \cellcolor[HTML]{EFEFEF}1    & \cellcolor[HTML]{EFEFEF}1  & 1  \\
20 & 25 & \multirow{-5}{*}{$\beta$} & \multirow{-5}{*}{4} & 0                & 0.02                         & 0                         & 0                 & 0                 & 0               & \cellcolor[HTML]{EFEFEF}1    & \cellcolor[HTML]{EFEFEF}1  & 1  \\ \hline
\end{tabular}
\end{table*}

\begin{table*}[t!]
\caption{Type~I error for CHASM for multivariate Pareto data.}
\label{tab:first_error_pareto}
\centering
\begin{tabular}{llllllllll}
\hline
no & d &  \textbf{$\Theta_{L-LOC}$}  & \textbf{$\Theta_{L-SCALE}$}  & \textbf{$\Theta_{BMID}$} & \textbf{$\Theta_{L-SKEW}$}  & \textbf{$\Theta_{L-KURT}$}  & \textbf{$\Theta_{CORR}$} & \textbf{$D_B$} & \textbf{$D_J$} \\ \hline
1  & 2       & 0      & 0.02    & 0            & 0          & 0          & 0             & 0.02               & 0.04         \\
2  & 3  &      0      & 0       & 0            & 0          & 0          & 0             & 0.02               & 0.02         \\
3  & 5  &      0      & 0       & 0            & 0          & 0          & 0             & 0.02               & 0.02         \\
4  & 10 &      0      & 0       & 0            & 0          & 0          & 0             & 0.02               & 0.02         \\
5  & 25     & 0      & 0       & 0            & 0          & 0          & 0             & 0.02               & 0.02 \\   \hline     
\end{tabular}
\end{table*}

In experiments~6 and~7 (see first two rows of Tab. \ref{tab:experiments_pareto}), we investigate changes in the parameter $\alpha$, which directly affect the scale of the distribution. Specifically, in experiment~6 the variance of the coordinates of multivariate Pareto distribution shifts from finite to infinite, whereas in experiment~7 such variance remains finite (but changes its magnitude). As summarized in Tab.~\ref{tab:power_pareto}, both the scale-sensitive criteria ($\Theta_{L-SCALE}, \Theta_{BMID}$) and the general criteria measuring distributional divergence ($D_J$ and $D_B$) consistently detect statistically significant changes in both scenarios.  

In experiments~8 and~9, the parameter $\beta$ is varied, which primarily influences the scale. In these cases as well, the general divergence-based criteria ($D_J$ and $D_B$) successfully indicate significant changes in the distribution. Analogously to the simulations made for multivariate Gaussian distribution, the last column ($\mathcal{R}$) of Tab.~\ref{tab:power_pareto} reports the fraction of simulation runs in which at least one of the criteria detected a change. Importantly, across the majority of experiments and for all considered dimensionalities $d$, the detection accuracy is perfect (statistical power equal to 1), meaning that no true change went undetected.  This implies that when $\alpha$ or $\beta$ parameters shift - whether affecting the overall scale or the tail behavior of the distribution - the employed methodology is highly reliable in identifying such distributional changes. This robustness holds even in high-dimensional settings, which are often encountered in software systems dealing with multivariate data streams.

We also assess the Type~I error of the statistical tests utilized within CHASM. To this end, we generated data of length $N = 400$ from the distribution $MP_{0.4, 1, d}$, which by construction contains no structural changes. The evaluation was performed over 100 Monte Carlo replications, where the CHASM procedure was applied by designating the first half of the sample as one segment and the second half as another. The corresponding results are presented in Tab.~\ref{tab:first_error_pareto}. For almost all examined dimensionalities $d$, no statistically significant changes were detected. Thus, regardless of the specific combination of criteria used within CHASM, the procedure yielded no false positives under these conditions.

Additionally, we also verify the computational complexity at runtime for selected tests performed within the CHASM procedure. The results are presented in Fig. \ref{fig:pareto_times}.

\begin{figure}[h!]
    \centering
    \includegraphics[width=0.9\linewidth]{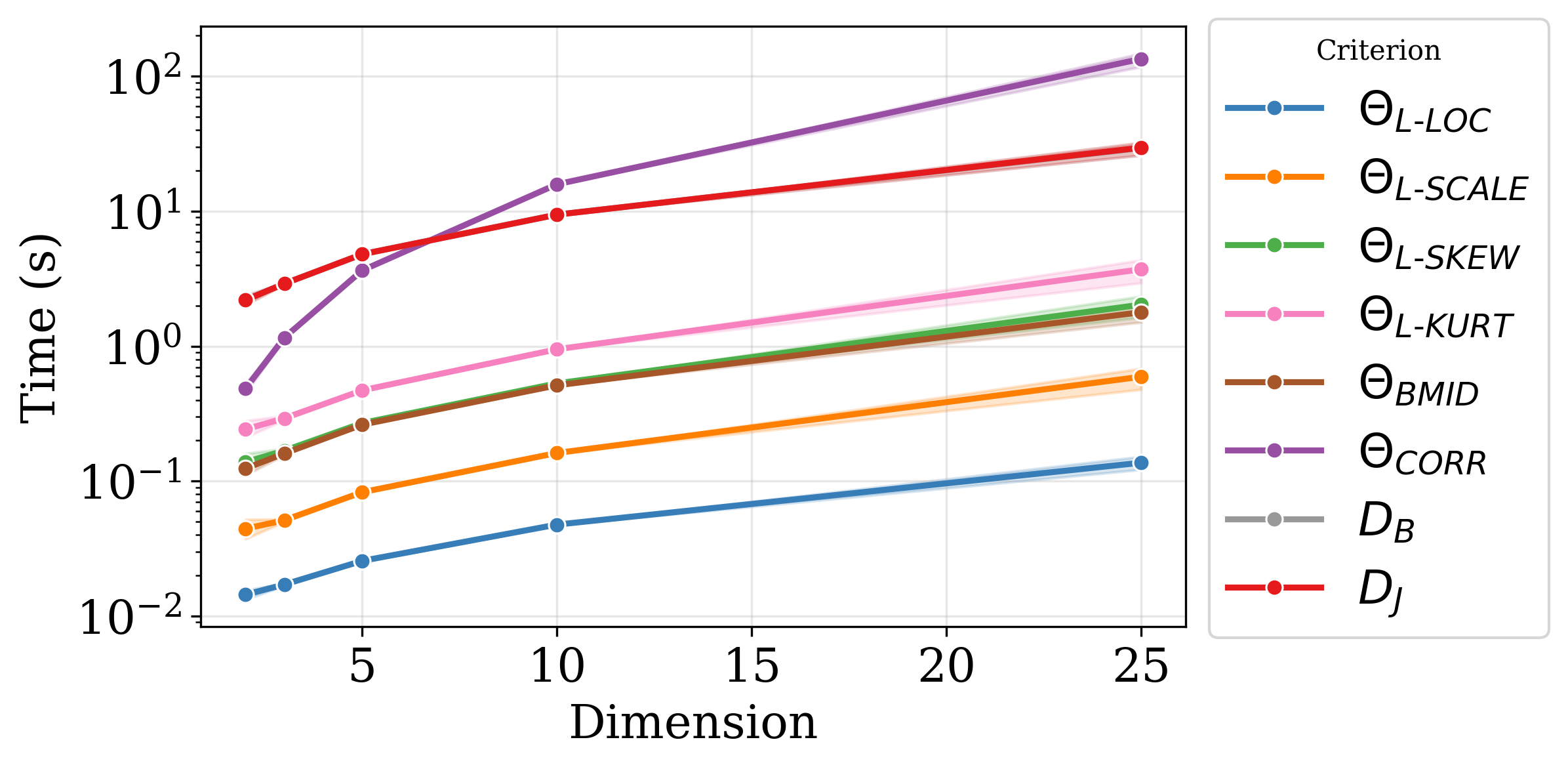}
    \caption{Runtime computational complexity for selected statistics used by CHASM for multivariate Pareto distributed data.}
    \label{fig:pareto_times}
\end{figure}

\newpage
\subsection{Evaluating CHASM over baseline segmentation}
\label{sec:multiple_cp}

In the following part, we evaluate the effect of applying CHASM to the assessment of the significance of change points obtained using KCPA described in Section \ref{sec:baseline_method} on multivariate Gaussian and Pareto distributions. The primary objective of this experiment is to assess how CHASM reduced the number of false positives detected by pure segmentation algorithm. For this purpose, we generate data from model~\eqref{eq:assumption_multiple_cp}, consisting of eight homogeneous segments.

In the first case, each segment follows a multivariate Gaussian distribution with the corresponding distribution parameters provided in Tab.~\ref{tab:experiments_gauss_multi}.

\begin{table}[h!]
\centering
\caption{Parameters of Gaussian distribution corresponding to the subsequent segments ($d=5$).}
\label{tab:experiments_gauss_multi}
\begin{tabular}{l|cccccccccc}
\hline
id   & 1        & 2              & 3        & 4              & 5        & 6        & 7        & 8            & 9        & 10           \\ \hline
$\bm{\mu}$           & $\mathbf{1}$ 
                & $\mathcal{V}$ 
                & $\mathbf{1}$ 
                & $\mathbf{2}$ 
                & $\mathbf{1}$ 
                & $\mathbf{1}$ 
                & $\mathbf{1}$ 
                & $\mathbf{1}$ 
                & $\mathbf{1}$ 
                & $\mathbf{1}$ \\
$\bm{\Sigma}$        & $I$ 
                & $I$ 
                & $I$ 
                & $I$ 
                & $I$ 
                & $2I$ 
                & $I$ 
                & $\bm{\Sigma}_1$ 
                & $I$ 
                & $\bm{\Sigma}_2$ \\ \hline
\end{tabular}
\end{table}
where:  
\begin{itemize}
    \item $\mathcal{V}$ denotes the vector $[2,1,1,1,1]$
    \item $\mathbf{1}$ denotes the vector $[1,1,1,1,1]$,  
    \item $\mathbf{2}$ denotes the vector $[2,2,2,2,2]$,  
    \item $I$ is the identity matrix,  
    \vspace{0.1cm}
    \item $\bm{\Sigma}_1 = \begin{bmatrix}
1 & 0.9 & 0.3 \\
0.9 & 1 & 0.2 \\
0.3 & 0.2 & 1 \\
\end{bmatrix}$ 
\vspace{0.1cm}
    \item $\bm{\Sigma}_2 = \begin{bmatrix}
1 & 0.5 & 0.3 \\
0.5 & 1 & 0.2 \\
0.3 & 0.2 & 1 \\
\end{bmatrix}$. 
\end{itemize} 
\vspace{0.1cm}
The KCPA method involves several tunable hyperparameters, including the choice of kernel, minimum segment size, jump, and penalty (detailed descriptions are available, e.g., in the Python package documentation related paper \cite{truong2020selective}). In our study, we set the minimum segment size to 200, the jump parameter to 1, and considered the following kernel and penalty configurations: kernel $\in \{\text{Gaussian}, \text{linear}, \text{Cauchy}\}$ and penalty $\in \{1, 2, 3\}$. For the Cauchy kernel, the imposed constraints were overly restrictive, typically resulting in the detection of only a single regime change; therefore, this configuration was excluded from the final visualizations. For each parameter configuration, dimension $d$, and simulation setting, we conducted 100 Monte Carlo replications. In each run, the data were generated, segmented using KCPA, and subsequently refined by applying CHASM, which excluded statistically insignificant change points. The results, expressed as the number of detected change points before and after applying CHASM, are presented in Fig.~\ref{fig:number_of_detected}. The dashed line in the figure denotes the true number of change points. The obtained results indicate that CHASM consistently achieves a lower false-positive rate, even under suboptimal parameter settings of KCPA.

\begin{figure}[h!]
    \centering
    \includegraphics[width=\linewidth]{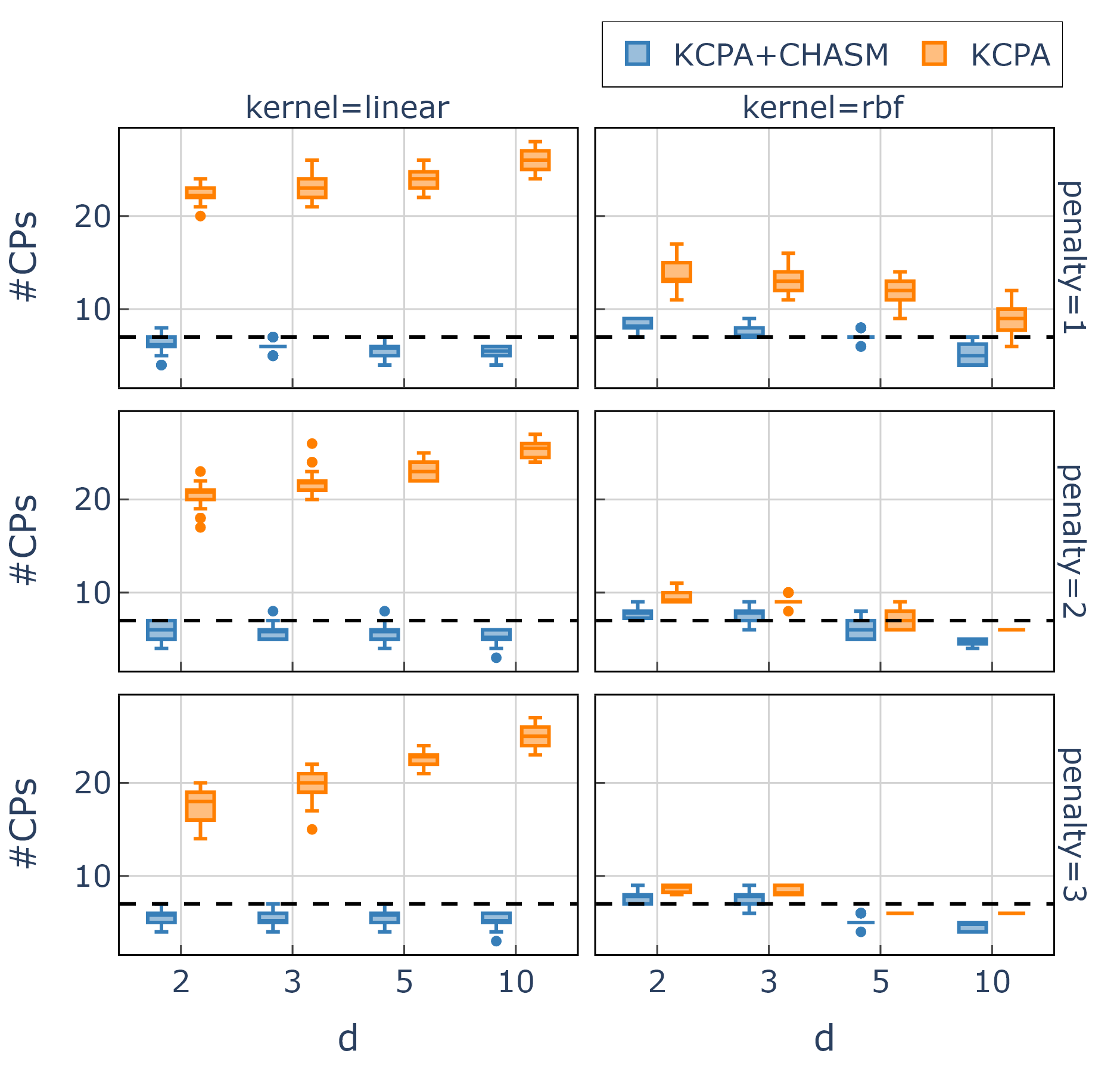}
    \caption{Number of detected change points for KCPA with and without CHASM - results for Gaussian distribution.}
    \label{fig:number_of_detected}
\end{figure}

In the second case, each segment follows the multivariate Pareto distribution, characterized by the density function given by Eq. \eqref{eq:pareto_pdf} , with the corresponding distribution parameters provided in Tab.~\ref{tab:experiments_pareto_multi}. 

\begin{table}[h!]
\centering
\caption{Parameters of Pareto distribution corresponding to the next segments.}
\label{tab:experiments_pareto_multi}
\begin{tabular}{l|llllllll}
\hline
id & 1   & 2   & 3   & 4   & 5   & 6   & 7   & 8   \\ \hline
$\alpha$   & 0.4 & 0.8 & 0.4 & 0.2 & 0.4 & 0.4 & 0.4 & 0.4 \\
$\beta$    & 1   & 1   & 1   & 1   & 1   & 10  & 0.2 & 1   \\ \hline
\end{tabular}
\end{table}

\begin{figure}[h!]
    \centering
    \includegraphics[width=\linewidth]{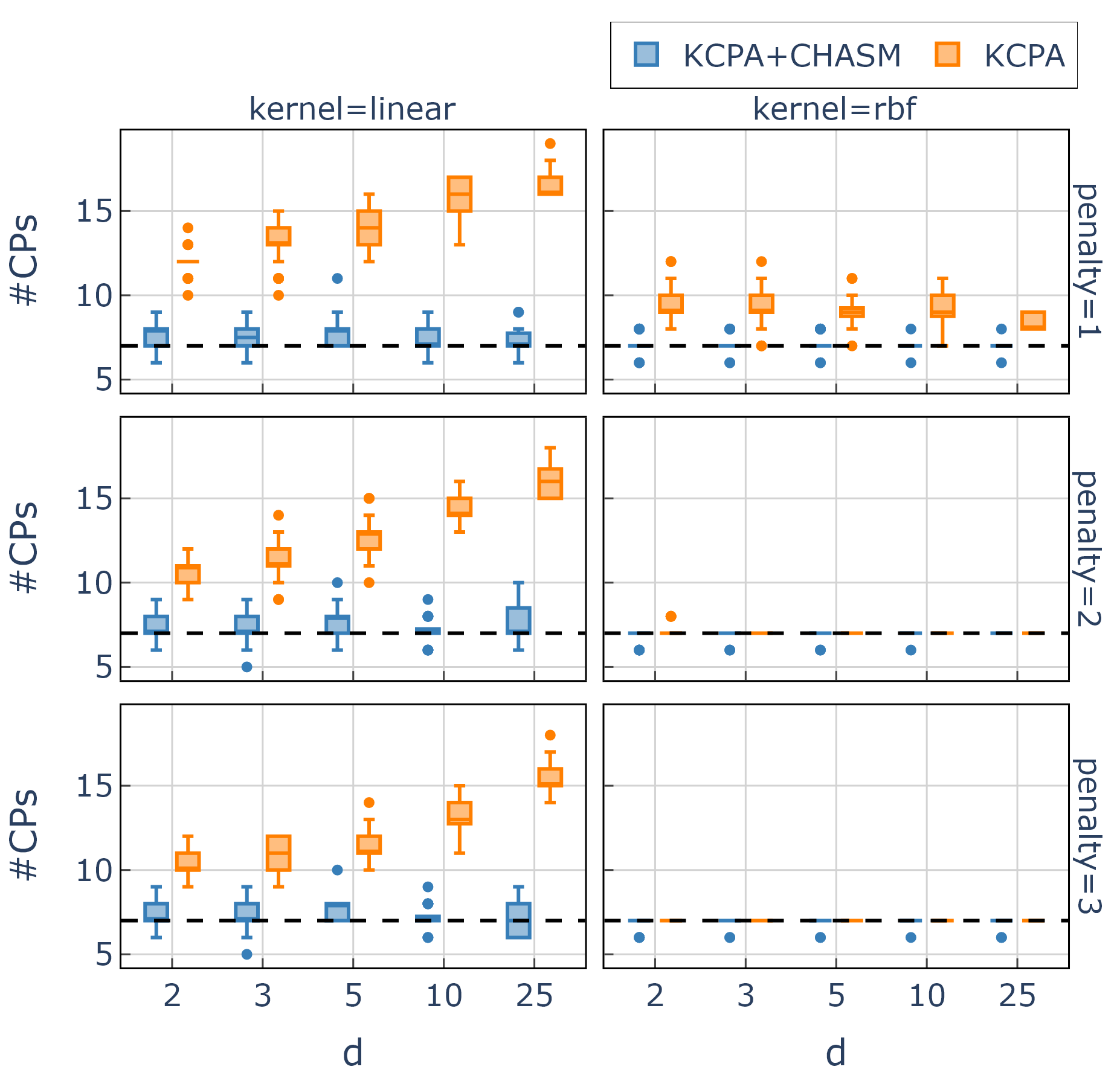}
    \caption{Number of detected change points for KCPA with and without CHASM - results for Pareto distribution.}
    \label{fig:pareto_sim_number_of_detected}
\end{figure}

Analogous results as for the Gaussian distribution - regarding the number of detected change points before and after applying CHASM for Pareto distributed data - are presented in Fig.~\ref{fig:pareto_sim_number_of_detected}. The dashed line in the figure denotes the true number of change points. 

Again, the results demonstrate that CHASM consistently reduces the number of false positives, even when KCPA is suboptimally parameterized. This is particularly relevant in practice, as segmentation algorithms often involve complex parameterization, and visual inspection of results is infeasible when working with large-scale datasets. Hence, CHASM provides a systematic and statistically grounded mechanism for improving segmentation reliability by filtering out spurious detections. The Mean Absolute Percentage Error (MAPE) for the number of detected change points obtained solely with \textsc{KCPA} is 46\%. However, when statistical significance is evaluated using \textsc{CHASM} and only significant changes are retained (authorization), the MAPE reduces to 2\%, representing a 23-fold improvement.

\section{Significance of change points detected in multidimensional industrial data}
\label{sec:real_data}

\subsection{Dataset 1: Industrial data of a reciprocating compressor}
\label{sec:real_data1}

The investigated data were acquired from a condition monitoring system of a reciprocating compressor in an oil and gas company. It is used to compress natural gas to be sent via a pipeline to a gas power plant. The compressor is powered by 2MW electric motor and has 4 cylinders, which are used to gradually increase the pressure of natural gas. {Multiple heterogeneous variables were observed to indicate damage, including (i) metrics derived from mechanical vibrations, and (ii) process variables, i.e., environmental descriptors of the complex system such as the pressure of specific components, temperature, and fluid level.

From a signal processing viewpoint, when damage of the compressor occurs, the characteristics of the data change. However, it is not entirely clear which data characteristics change in case of damage, as the signal contains various components. Therefore, it is necessary to use a universal approach that allows for the automatic identification of change points and their significance.

The raw data were originally sampled at 15-second intervals and subsequently preprocessed by temporal aggregation to a 1-minute resolution that gives in total 7694 timesteps. Representative process variables used in the analysis are presented in Fig.~\ref{fig:process_data}, while vibration metrics, including peak-to-peak (PP), root mean square (RMS), and vibration velocity root mean square (VRMS) measured by sensors mounted in the horizontal and vertical orientations on a single cylinder are shown in Fig.~\ref{fig:trends_data}. The signals corresponding to the remaining three cylinders exhibit analogous characteristics and are therefore not included in this work.} The compressor developed a damage, which was detected by an increase in mechanical vibration {(see, 1400th sample in Fig.~\ref{fig:trends_data})}. Further damage occured at sample 3533, which can be seen in the data by sudden increase of the Cylinder I VRMS parameter. The next change was at the sample 4000, when the fault was detected. The detection occurred at a stage when vibrations were already elevated for a couple of weeks. After the detection the compressor was repaired (note, that the data when compressor was offline are not in the dataset, as the monitoring system was inactive). Finally, the 4th change point occured around sample 5000, when PP parameters of both vibration channels changed its mean value, responding to the level control mode.

\begin{figure}[h!]
	\centering
	\includegraphics[width=\linewidth]{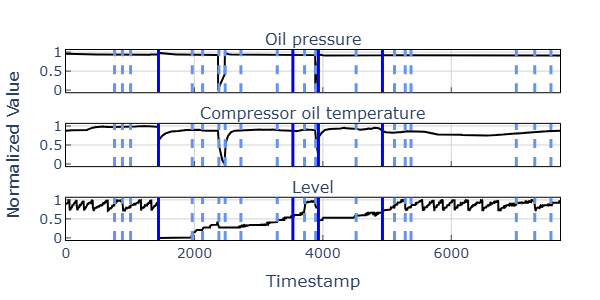}
	\caption{Process data segmentation with significance of changes assessment for Dataset 1. Blue lines correspond to change points detected by some segmentation algorithm (dashed lines correspond to false positives and solid lines are true change point).}
	\label{fig:process_data}
\end{figure}

\begin{figure}[h!]
    \centering
    \includegraphics[width=\linewidth]{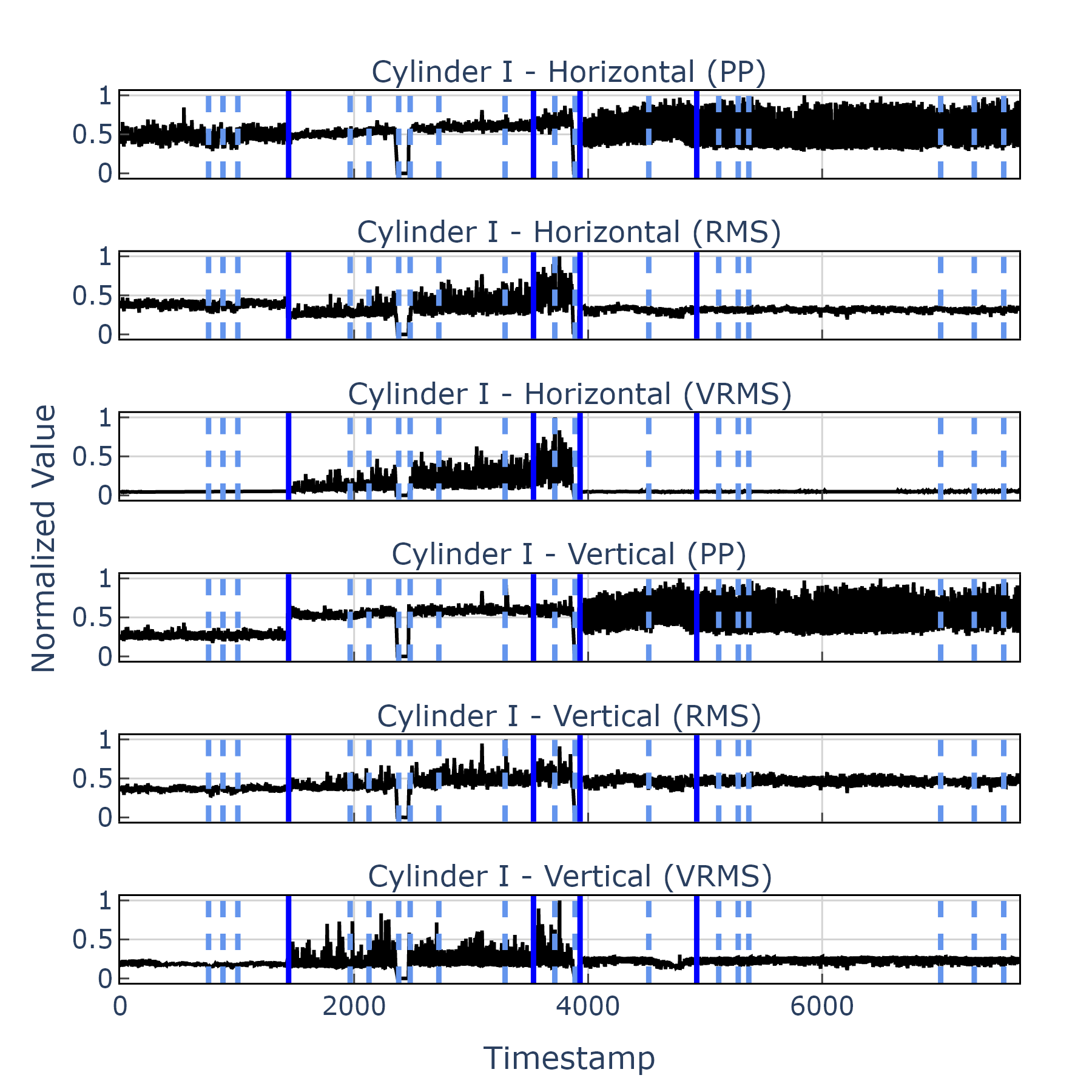}
    \caption{Significant and insignificant change points detected for process data in comparison with vibration metrics indicating machine failure for Dataset 1. Blue lines correspond to change points detected by some segmentation algorithm (dashed lines correspond to false positives and solid lines are true change points).}
    \label{fig:trends_data}
\end{figure}

As presented in  {Fig.~\ref{fig:trends_data}}, the data cover timespan of different machine behavior. First, the compressor worked properly and was in good condition. Next, the health begun to deteriorate, but was not yet detected by the condition monitoring system (CMS). When CMS detected the damage, the machine was repaired and the final part of the data present operation afterwards.

\begin{table*}[b!]
\centering
\caption{Explainability analysis for change point $n_1=1400$  for Dataset 1.}
\label{tab:explainability}
\begin{tabular}{l|l|l|l|l|l|l|l|l}
\hline
\textbf{cp}             & \textbf{variable} & \textbf{$\Theta_{L-SCALE}$}             & \textbf{$\Theta_{L-LOC}$}              & \textbf{$\Theta_{BMID}$}        & \textbf{$\Theta_{L-SKEW}$}          & \textbf{$\Theta_{L-KURT}$}          & \textbf{$D_B$}  & \textbf{$D_J$}        \\ \hline
                        & v1                &                               & \cellcolor[HTML]{8DCA7D}0.149 & \cellcolor[HTML]{FED17F}1.926 &                               &                               & \cellcolor[HTML]{FFDD82}1.292 & \cellcolor[HTML]{FAE983}0.521 \\
                        & v2                & \cellcolor[HTML]{6BC07B}0.034 & \cellcolor[HTML]{81C67C}0.110 & \cellcolor[HTML]{FED781}1.609 &                               &                               & \cellcolor[HTML]{FFDD82}1.269 & \cellcolor[HTML]{FFEB84}0.538 \\
                        & v3                &                               & \cellcolor[HTML]{AAD27F}0.250 & \cellcolor[HTML]{FFDB81}1.392 &                               &                               & \cellcolor[HTML]{FFE283}1.045 & \cellcolor[HTML]{B0D47F}0.271 \\
                        & v4                &                               &                               & \cellcolor[HTML]{F7E883}0.511 &                               & \cellcolor[HTML]{9FCF7E}0.211 & \cellcolor[HTML]{FFE884}0.721 & \cellcolor[HTML]{91CB7D}0.164 \\
                        & v5                & \cellcolor[HTML]{63BE7B}0.005 &                               &                               &                               &                               & \cellcolor[HTML]{FDC57D}2.517 & \cellcolor[HTML]{FDC47D}2.578 \\
                        & v6                &                               &                               &                               &                               &                               &                               &                               \\
                        & v7                &                               &                               &                               &                               &                               &                               &                               \\
                        & v8                &                               &                               & \cellcolor[HTML]{64BE7B}0.011 &                               &                               &                               & \cellcolor[HTML]{69BF7B}0.029 \\
                        & v9                &                               & \cellcolor[HTML]{F8E983}0.514 &                               &                               &                               & \cellcolor[HTML]{9DCE7E}0.206 & \cellcolor[HTML]{A2D07E}0.223 \\
                        & v10               &                               &                               &                               &                               &                               & \cellcolor[HTML]{FFDF82}1.207 & \cellcolor[HTML]{FCA377}4.271 \\
                        & v11               &                               &                               &                               &                               &                               &                               &                               \\
                        & v12               &                               &                               &                               &                               &                               &                               &                               \\
                        & v13               &                               &                               &                               &                               &                               &                               &                               \\
                        & v14               &                               &                               & \cellcolor[HTML]{F8696B}7.243 & \cellcolor[HTML]{FECD7F}2.135 &                               & \cellcolor[HTML]{FDBB7B}3.019 & \cellcolor[HTML]{8BC97D}0.142 \\
                        & v15               &                               &                               & \cellcolor[HTML]{FED07F}1.939 &                               & \cellcolor[HTML]{ABD27F}0.252 &                               &                               \\
                        & v16               &                               &                               & \cellcolor[HTML]{FECD7F}2.112 &                               &                               & \cellcolor[HTML]{FFDF82}1.165 & \cellcolor[HTML]{FCEA83}0.530 \\
                        & v17               &                               &                               & \cellcolor[HTML]{FED580}1.724 &                               &                               & \cellcolor[HTML]{FFE683}0.824 & \cellcolor[HTML]{BDD880}0.315 \\
\multirow{-18}{*}{1440} & v18               &                               &                               & \cellcolor[HTML]{FED881}1.545 &                               & \cellcolor[HTML]{B4D57F}0.284 &                               &                               \\ \hline 
\end{tabular}
\end{table*}

If the monitoring or control system was equipped with a method to detect such an early changes in a multidimensional process data, it could inform the maintenance department and repair can be prepared in advance. Sometimes only a change in operational regime is sufficient to decrease a damage progress rate. {Based on specific vibration metrics, it is possible to identify abnormal operating conditions of individual system components. However, this approach is associated with high costs, including the installation of numerous sensors and the memory resources required for continuous processing of high-volume vibration data. As an alternative, in addition to vibration measurement software, the sensors are employed that capture fundamental process variables such as temperature, pressure, and fluid levels associated with specific components of the system. The primary objective of this study is to evaluate whether operational anomalies can be detected through the analysis of significant variations in these process variables.} This is accomplished using the methodology described in Section~\ref{sec:proposed_approach}.

{The compressor is subject of a continuous analysis by an expert team, which uses in-house analysis methods. According to the information we were able to obtain, these analysis methods are a set of multidimensional analyses, using extensive amount of expert knowledge. Based on this input, three process variables were identified as valuable for damages analyses. The selected variables are presented in Fig.~\ref{fig:process_data}. The resulting time series was segmented using the KCPA method, employing a Gaussian kernel. 
Segmentation was performed with a minimum segment length of 30 samples, a step size of 1, and a penalty parameter of 15. All parameter values were determined empirically. A detailed description of the method and parameter selection is provided in~\cite{truong2020selective}.}

{In the next step, CHASM was used to assess the statistical significance of the detected change points. The used parameters are: $\omega=200$, $r=1000$ (default) and $c=95\%$ (default). The value of confidence level $c$ is a commonly used value in statistics, $r$ denotes the resampling number, so it should be reasonably large, and $\omega$ was chosen due to the characteristics of the data and the underlying statistics used as \textit{basic statistics}. Fig.~\ref{fig:process_data} illustrates the segmentation of the process data along with the assessment of the significance of change, where blue lines mark the detected change points (solid lines indicate true changes, while dashed lines represent false positives). The detected change points (both true positives and false positives) are also indicated in Fig.~\ref{fig:trends_data}, where the vibration metrics are presented. As observed, the detected changes exhibit a strong correspondence with the actual variations identified from the vibration data metrics.} 

Furthermore, this work seeks to investigate the relationship between considered variables and component failures, thus providing information on the explainability of failure mechanisms. This is accomplished using the methodology described in Section~\ref{sec:proposed_approach2}. We begin by examining the significant change points detected in the last paragraph, denoted as $n_1$, $n_2$, $n_3$, and $n_4$. For all 18 available process variables, these change points are localized within the signal. Subsequently, for each univariate trajectory, each change point ($n_1$--$n_4$), and each criterion (among the considered set), we assess the statistical significance of the detected changes.

Suppose criterion $L$ indicates a change for variable $v_1$. Our objective is to determine the extent to which variable $v_1$ deviates from its baseline with respect to criterion $L$. For this purpose, we define the following measure of the \emph{magnitude of change}:  

\begin{align}
    \text{MoC} = \frac{|S^L_B - S^L_T|}{|S_B|},
    \label{eq:moc}
\end{align}
where $S^L_B$ denotes the boundary value of the significance interval, and $S^L_T$ is the test statistic obtained for criterion $L$. The results for the first detected change point ($n_1$), corresponding to the onset of machine failure, are reported in Tab.~\ref{tab:explainability}. Tables for the remaining change points are provided \ref{app:simulated_data1}. Empty entries in Tab.~\ref{tab:explainability} indicate the absence of significant changes for a given variable under the respective criterion.  
As can be observed, a particularly influential factor potentially linked to the machine failure is variable $v_{14}$, which exhibits a substantial scale change (criterion $\Theta_{BMID}$) at the detected change point. Thus, the explanation of the failure should primarily be sought in the component associated with this variable. Nevertheless, other variables, namely $v_1$--$v_4$ and $v_{15}$--$v_{18}$, also should be investigated in the future, since they display considerable scale changes as well. Interestingly, for variables $v_5$ and $v_{10}$, other distributional characteristics underwent significant shifts, as reflected in the MoC values for criterion $D_J$.


\subsection{Dataset 2: Vibration data from motorbike}
\label{sec:real_data2}

The second example of real data comes from an experiment that is based on vibration measurements collected with a smartphone equipped with motion sensors \cite{witulska2026vibration}. The data were acquired using the Physics Toolbox Sensor Suite application for Android. The smartphone sensors recorded motion along three axes, providing measurements from a G-Force Meter, a linear accelerometer, and a gyroscope. The maximum sampling frequency of the sensors was 100~Hz. For more details about dataset and experiment, see \cite{witulska2026vibration, witulska2022identification}.

\begin{figure}[h!]
    \centering
    \includegraphics[width=\linewidth]{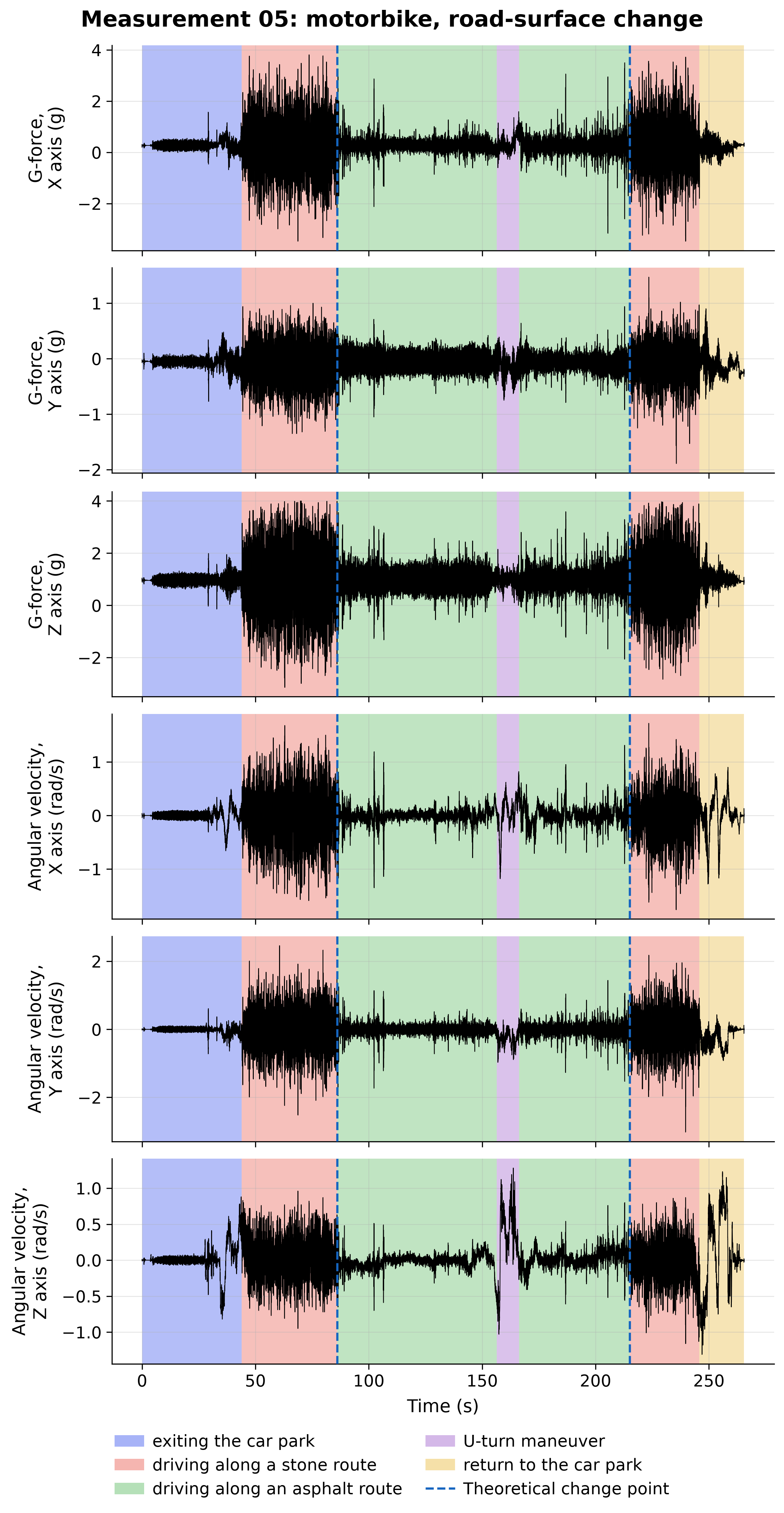}
    \caption{Motorbike vibration signals (G-force and angular velocity, three axes) recorded while driving over cobblestone and asphalt. Coloured bands denote labeled route segments; blue dashed lines mark the theoretical change points at 86 s and 215 s (midpoints of the observer-declared intervals).}
    \label{fig:vibro_motor}
\end{figure}

In this study, we use only measurement no.~5 from the original measurement set. The experiment was performed on a motorbike and lasted $265.49$~s. Two sensors were used: the G-Force Meter and the gyroscope. Each sensor records measurements along three axes; thus, a total of six signals are obtained (see, Fig.~\ref{fig:vibro_motor}). During the experiment, the motorbike travelled over surfaces with different characteristics. The change in the road surface was therefore the main factor responsible for changes in the recorded vibration pattern. In particular, two changes of surface were observed during the measurement. The corresponding time intervals were $[85,87]$~s and $[214,216]$~s. Since the exact transition time was determined by visual observation during the experiment, the midpoints of these intervals, i.e., $86$~s and $215$~s, were used as reference change-point locations. When analyzing the MoC indicator defined by Eq.~(\ref{eq:moc}), we evaluated the CHASM procedure for these true change points identified based on the methodologies presented  in articles \cite{witulska2026vibration, witulska2022identification}. These points correspond to regime changes associated with changes in the signal scale resulting from changes in the road surface on which the vehicle was traveling. Although changes associated with vehicle maneuvers, such as turning (e.g., entering a parking area or performing a U-turn), could also be considered, we assumed that the significance of the identified change points should be assessed according to the criterion proposed in \cite{witulska2022identification}.

For each univariate trajectory, each change point, and each criterion, the statistical significance of the detected change is assessed by CHASM with window length \(\omega=200\) observations, \(r=1000\) resamples, and confidence level \(c=95\%\). The results for \(n_1\) and \(n_2\) are reported in Tabs.~\ref{tab:explainability_n1} and~\ref{tab:explainability_n2}.

\begin{table*}[!t]
\centering
\caption{Explainability analysis for change point \(n_1=86\,\mathrm{s}\) (sample \(17298\)) for Dataset 2.}
\label{tab:explainability_n1}
\begin{tabular}{l|l|l|l|l|l|l|l|l}
\hline
\textbf{cp} & \textbf{variable} & \textbf{\(\Theta_{L\textrm{-SCALE}}\)} & \textbf{\(\Theta_{L\textrm{-LOC}}\)} & \textbf{\(\Theta_{BMID}\)} & \textbf{\(\Theta_{L\textrm{-SKEW}}\)} & \textbf{\(\Theta_{L\textrm{-KURT}}\)} & \textbf{\(D_B\)} & \textbf{\(D_J\)} \\
\hline
 & gforce\_x
 & \cellcolor[HTML]{F8E983}0.450
 &
 & \cellcolor[HTML]{FED17F}0.590
 &
 &
 & \cellcolor[HTML]{FDBB7B}3.595
 & \cellcolor[HTML]{F8696B}5.595 \\
 & gforce\_y
 & \cellcolor[HTML]{63BE7B}0.001
 &
 &
 &
 &
 &
 & \cellcolor[HTML]{FFDD82}1.057 \\
 & gforce\_z
 & \cellcolor[HTML]{AAD27F}0.162
 & \cellcolor[HTML]{F8E983}0.446
 &
 &
 &
 & \cellcolor[HTML]{FDC57D}2.518
 & \cellcolor[HTML]{FDBB7B}3.059 \\
 & ang\_vel\_x
 & \cellcolor[HTML]{AAD27F}0.151
 &
 &
 &
 &
 & \cellcolor[HTML]{FFDD82}1.677
 & \cellcolor[HTML]{FDBB7B}3.601 \\
 & ang\_vel\_y
 & \cellcolor[HTML]{F7E883}0.289
 &
 & \cellcolor[HTML]{8DCA7D}0.082
 &
 &
 & \cellcolor[HTML]{FDC57D}2.395
 & \cellcolor[HTML]{F8696B}4.191 \\
\multirow{-6}{*}{17298} & ang\_vel\_z
 & \cellcolor[HTML]{8DCA7D}0.081
 &
 &
 &
 &
 & \cellcolor[HTML]{FDC57D}2.916
 & \cellcolor[HTML]{F8696B}4.351 \\
\hline
\end{tabular}
\end{table*}

\begin{table*}[!t]
\centering
\caption{Explainability analysis for change point \(n_2=215\,\mathrm{s}\) (sample \(43005\)) for Dataset 2.}
\label{tab:explainability_n2}
\begin{tabular}{l|l|l|l|l|l|l|l|l}
\hline
\textbf{cp} & \textbf{variable} & \textbf{\(\Theta_{L\textrm{-SCALE}}\)} & \textbf{\(\Theta_{L\textrm{-LOC}}\)} & \textbf{\(\Theta_{BMID}\)} & \textbf{\(\Theta_{L\textrm{-SKEW}}\)} & \textbf{\(\Theta_{L\textrm{-KURT}}\)} & \textbf{\(D_B\)} & \textbf{\(D_J\)} \\
\hline
 & gforce\_x
 &
 &
 &
 &
 &
 &
 & \\
 & gforce\_y
 & \cellcolor[HTML]{8DCA7D}0.068
 &
 & \cellcolor[HTML]{AAD27F}0.151
 &
 &
 &
 & \cellcolor[HTML]{F8E983}0.349 \\
 & gforce\_z
 &
 &
 & \cellcolor[HTML]{F8E983}0.318
 &
 &
 &
 & \\
 & ang\_vel\_x
 & \cellcolor[HTML]{AAD27F}0.105
 &
 & \cellcolor[HTML]{AAD27F}0.188
 &
 &
 &
 & \\
 & ang\_vel\_y
 &
 &
 & \cellcolor[HTML]{F8E983}0.394
 &
 & \cellcolor[HTML]{8DCA7D}0.054
 &
 & \cellcolor[HTML]{63BE7B}0.019 \\
\multirow{-6}{*}{43005} & ang\_vel\_z
 &
 &
 & \cellcolor[HTML]{AAD27F}0.118
 &
 &
 &
 & \\
\hline
\end{tabular}
\end{table*}

As can be observed in Tab.~\ref{tab:explainability_n1}, a particularly influential factor associated with the first surface transition is variable \texttt{gforce\_x}, which exhibits both a pronounced scale change ($MoC$ for \(\Theta_{L\textrm{-SCALE}}=0.450\), ($MoC$ for $\Theta_{BMID}=0.590)$) and the largest $MoC$ for distributional distances among all channels (\(D_B=3.595\), \(D_J=5.595\)). The vertical acceleration \texttt{gforce\_z} likewise undergoes a significant location shift (\(\Theta_{L\textrm{-LOC}}=0.446\)) together with substantial scale and divergence effects. Thus, the explanation of the cobblestone-to-asphalt change should primarily be sought in the longitudinal and vertical g-force measurements.

Nevertheless, the three gyroscope channels should also be investigated, since they display considerable Jeffrey and Bhattacharyya distances. In particular, \texttt{ang\_vel\_y} and \texttt{ang\_vel\_z} attain \(D_J=4.191\) and \(D_J=4.351\), respectively, indicating that the change is not confined to a single sensor axis. By contrast, \texttt{gforce\_y} is only weakly implicated: the L-scale effect is at the detection threshold (\(\Theta_{L\textrm{-SCALE}}=0.001\)), and the remaining significant contribution is a moderate divergence (\(D_J=1.057\)).

Interestingly, neither L-skewness nor L-kurtosis is significant for any variable at \(n_1\). The break is therefore consistent with a change in scale and in the overall distribution, rather than with a modification of higher-order shape characteristics. This observation is in agreement with the observer labeling of the recording as a variance-change experiment.

The second transition, reported in Tab.~\ref{tab:explainability_n2}, is weaker and more heterogeneous. The largest MoC values are obtained for the robust scale of \texttt{ang\_vel\_y} (\(\Theta_{BMID}=0.394\)) and \texttt{gforce\_z} (\(\Theta_{BMID}=0.318\)), with additional significant scale effects on \texttt{gforce\_y}, \texttt{ang\_vel\_x}, and \texttt{ang\_vel\_z}. Thus, the return to cobblestone is expressed mainly as a change in dispersion. Distributional distances are much smaller than at \(n_1\): \(D_B\) is nowhere significant, and \(D_J\) is relevant only for \texttt{gforce\_y} (\(0.349\)) and, marginally, for \texttt{ang\_vel\_y} (\(0.019\)). No significant change is detected for \texttt{gforce\_x}, despite its dominant role at \(n_1\). Location and skewness remain insignificant, while a single weak kurtosis effect appears for \texttt{ang\_vel\_y} (\(\Theta_{L\textrm{-KURT}}=0.054\)).

Overall, the first change point is the more informative event for explainability: it is dominated by scale-aware L-statistics and by divergence criteria, with \texttt{gforce\_x} as the leading channel and with supporting evidence from \texttt{gforce\_z} and the gyroscope. The second change point is detected primarily through the robust scale criterion \(\Theta_{BMID}\) and does not reproduce the same channel ranking. This asymmetry indicates that the two surface transitions, although related, are not statistically equivalent when assessed in local windows around the annotated instants.

\section{Conclusions}
\label{sec:conclusion}

Detecting and analyzing change points in multivariate time series data is a critical task across various domains, including finance, engineering, and environmental science, as it determines the mid- to long-term security and reliability of automated and zero-touch systems and processes \cite{False_Alarms1,PatSnap_False_Alarms}. A change point signifies a shift in the underlying statistical properties of a system, such as a change in mean, variance, or correlation structure. Although there are numerous algorithms for segmenting the data and identifying these points, a relevant challenge remains: assessing the statistical significance of a detected change and its interpretation. Known segmentation techniques, particularly those based on heuristic or machine learning approaches, do not provide a measure of confidence, leading to a high rate of false alarms. These false alarms can result, e.g. in considerable financial loss and a lack of trust in automated monitoring systems. The problem becomes more complicated when the multivariate data exhibit impulsive (non-Gaussian heavy-tailed) behavior, which is a natural phenomenon in the real environment. {Finally, unlike many algorithms described in the literature, the approach proposed in this paper does not aim at evaluating change point detection methods under the assumption of known ground truth. Instead, it focuses on assessing the statistical significance of detected change points, which shifts the perspective from method evaluation toward increasing the certainty of results.}

In this article, we have introduced the non-parametric algorithm for evaluating the significance of the detected change points for multivariate non-Gaussian data - CHASM. Orchestrated with any change point detecttion algorithm, useful for data segmentation, CHASM operates in two complementary steps: the first step enables the assessment of statistical significance of detected changes in multivariate data (authorization), while the second step supports root-cause analysis, providing insights into the underlying factors driving these changes.} 

The primary motivation for our research comes from condition monitoring, where operating conditions of mechanical systems (e.g., industrial machinery) can change temporarily. Change in conditions naturally affects the data pattern that describe the operational parameters of such systems. Consequently, information that data have changed their probabilistic properties may indicate, e.g. a progressive damage to these systems. However, the proposed algorithm can also be considered as a universal method in other areas of interest, such as financial markets, where a change in the properties of the data described (e.g., stock indices) reflects market dynamics, and forecasting based on a homogeneous model can lead to erroneous conclusions. 

{The presented experimental studies confirm the effectiveness of the proposed method. The \textit{trustworthy factor} for simulation tests for {the multivariate Gaussian distribution (Tab.~\ref{tab:power_gauss}) averages 93.88\%}, whereas for the Pareto distribution it reaches 100\% (Tab.~\ref{tab:power_pareto}). This indicates that, regardless of the data dimensionality or the assumed distribution, the procedure is highly effective in determining which changes are significant and which are not.} 

Overall, the results obtained for multivariate Pareto data (Section~\ref{sec:pareto}) and contaminated multivariate Gauss data (\ref{app:cont_gauss}) demonstrate that, when appropriately selected and combined with a suitable observation window, the proposed criteria can provide reliable regime-change detection while maintaining a low false-positive rate, even under challenging contaminated-data conditions.

{Analyses of the results for datasets without significant changes (Tab.~\ref{tab:first_error_gauss} and Tab.~\ref{tab:first_error_pareto}) demonstrate that \textsc{CHASM} performs very well also for task of assessing false change points as not significant. Type~I error (proportion of false positives indicated as significant) is from 0\% to 4\%. The same conclusion can be drawn from the evaluation of change point significance in simulations involving multiple changes (Fig.~\ref{fig:pareto_sim_number_of_detected}). The MAPE computed for the number of detected change points using only \textsc{KCPA} amounts to 46\%, whereas after assessing significance with \textsc{CHASM} and retaining only significant changes (authorization), the MAPE decreases to 2\%--that is, a 23-fold improvement. }

\section*{Acknowledgment}
The work  is supported by National Center of Science under Weave-Unisono project \\No. 2025/07/Y/ST8/00070 ”Advanced signal processing techniques for cyclostationary modelling in Gaussian and non-Gaussian noisy environment - detection of cyclic sources, estimation, optimisation of algorithms and validation in the context of fault identification”.

\bibliographystyle{elsarticle-num} 
\bibliography{mybibliography}


\appendix
\label{sec:appendix}

\section{Simulated data examples} \label{app:simulated_data}

Fig. \ref{fig:gauss1} presents the realization of the two-dimensional Gaussian vector (described by \eqref{eq:gauss_multidim_pdf}) with $\bm{\mu} = [10, 0]$ and $\bm{\Sigma} = \begin{bmatrix}
2 & 0.9\sqrt{2} \\
0.9\sqrt{2} & 2
\end{bmatrix}$. 

The upper panels show the first component of the two-dimensional vector, while the bottom panels show the second components. 

\begin{figure}[h!]
    \centering
    \includegraphics[width=0.45\textwidth]{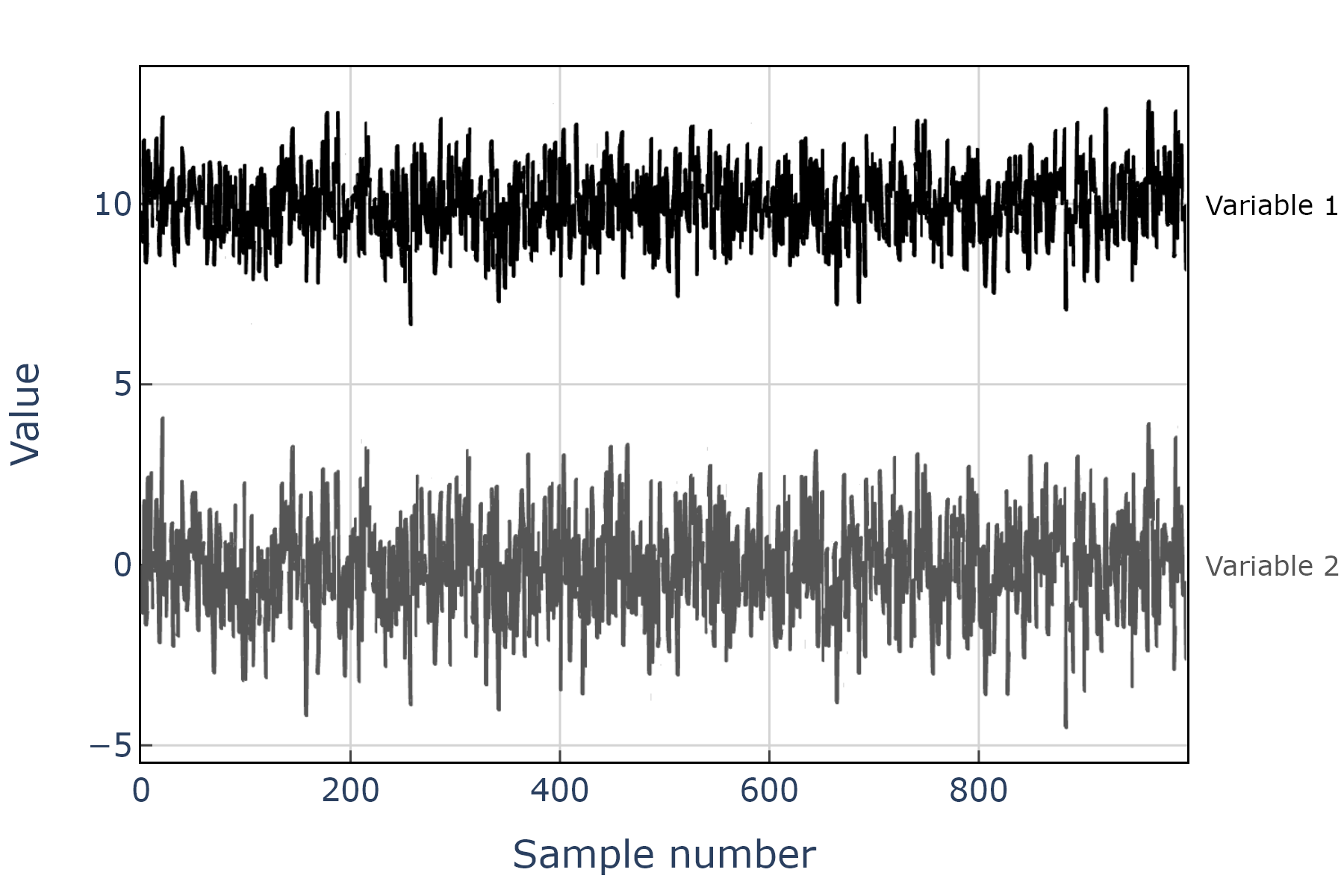}
    \caption{Two-dimensional Gaussian vector.}
    \label{fig:gauss1}
\end{figure}

The panels of Figs. \ref{fig:pareto1} and \ref{fig:pareto2} present the realizations of the two-dimensional Pareto vectors (described by \eqref{eq:pareto_pdf}). The upper panels show the first component of the two-dimensional vector, while the bottom panels show the second components. 

Fig.~\ref{fig:pareto1} shows the realization of the random vector with finite variance $\bm{X} \sim MP(\alpha=0.2, \beta=1, d=2) + [9, 0]$ (we added the mean shift only for the purpose of clear visualization). 

\begin{figure}[h!]
    \centering
    \includegraphics[width=0.45\textwidth]{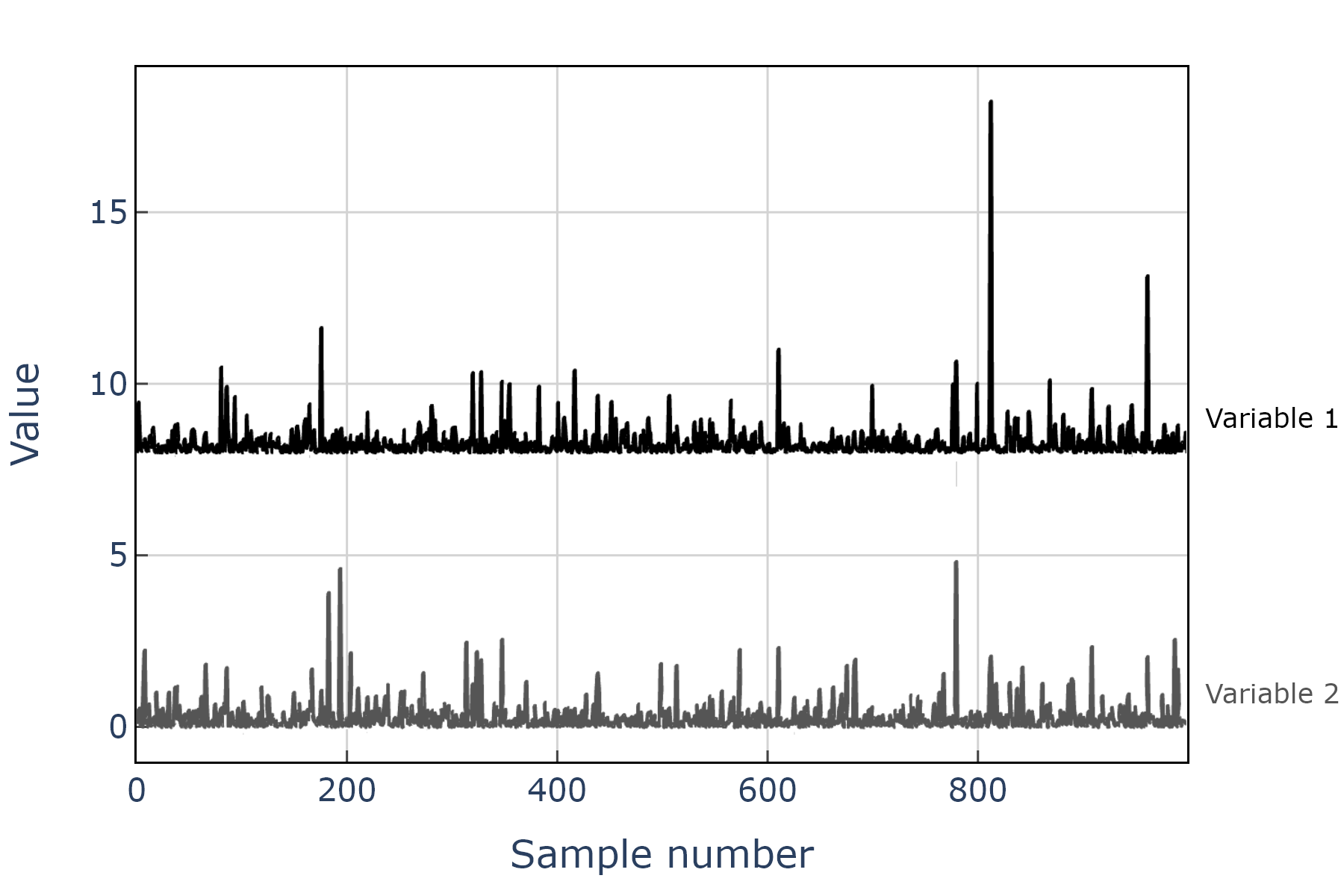}
    \caption{Two-dimensional Pareto vector with finite variance.}
    \label{fig:pareto1}
\end{figure}

\begin{figure}[h!]
    \centering
    \includegraphics[width=0.45\textwidth]{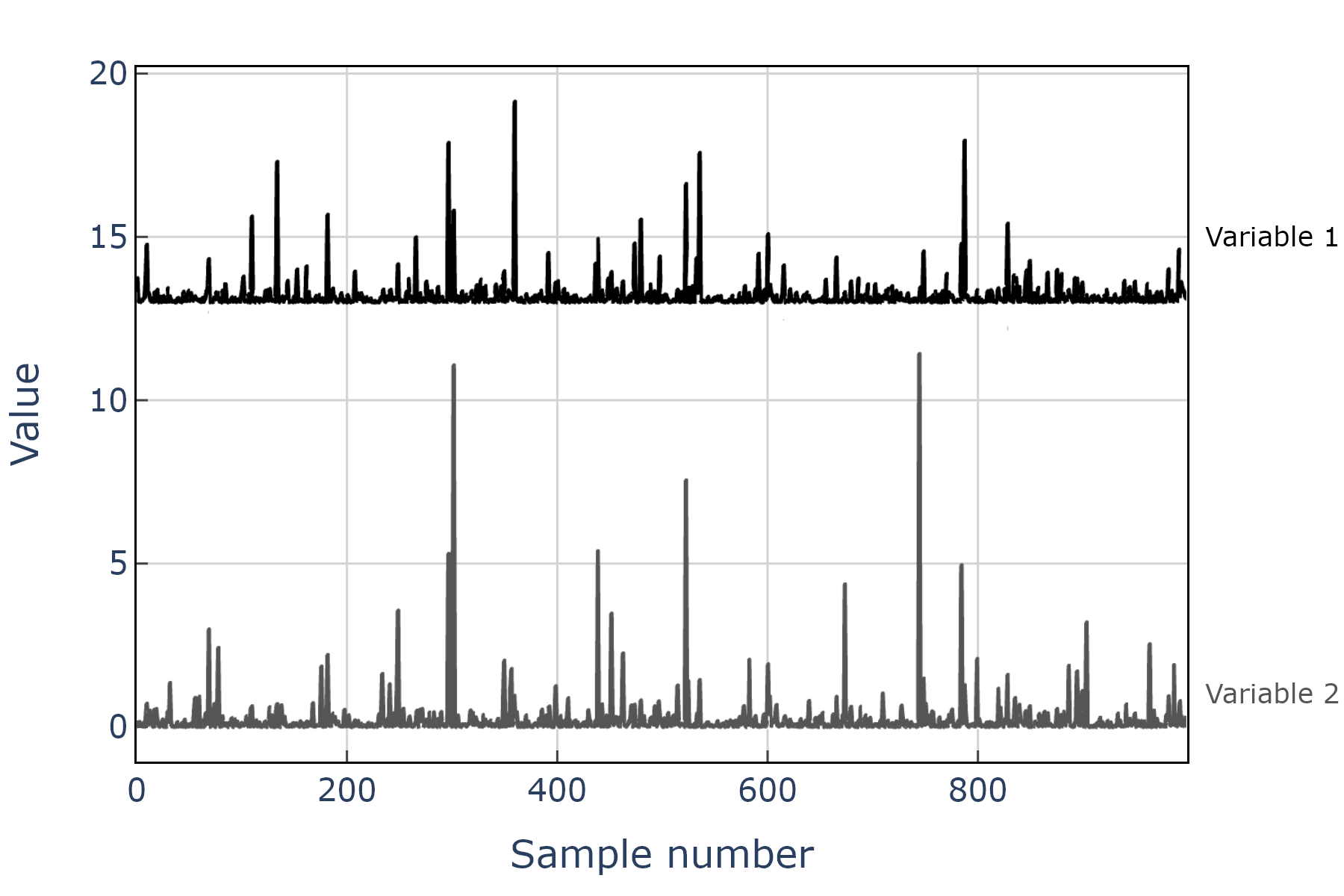}
    \caption{Two-dimensional Pareto vector with infinite variance.}
    \label{fig:pareto2}
\end{figure}

Otherwise, Fig.~\ref{fig:pareto2} shows the realization of the random vector with infinite variance $\bm{X} \sim MP(\alpha=0.8, \beta=10, d=2) + [14, 0]$ (we added the mean shift only for the purpose of clear visualization).

\section{Proposed CHASM vs likelihood ratio test}
\label{app:likelihood}
In this section, we complete the comparative analysis of the proposed CHASM method against a likelihood-ratio (LR) test \cite{neyman1992problem}, a classical parametric procedure for assessing if the compared samples are homogeneous or not.

The parametric LR procedure is applied to a pair of consecutive segments \(X^{(1)},X^{(2)}\in\mathbb{R}^{n\times d}\), each of length \(n=200\). Testing is performed coordinate-wise. For every dimension \(j=1,\ldots,d\), the hypotheses
\begin{equation}
H_{0,j}:\; \theta_j^{(1)}=\theta_j^{(2)}
\qquad\text{versus}\qquad
H_{1,j}:\; \theta_j^{(1)}\neq\theta_j^{(2)}
\end{equation}
are considered, where \(\theta_j^{(k)}\) denotes the parameter of the assumed univariate model on the \(j\)th coordinate of segment \(k\). Let \(\ell(\theta;x)\) denote the corresponding log-likelihood. The LR statistic is
\begin{equation}
\Lambda_j
=
2\Bigl(
\ell\bigl(\hat\theta_j^{(1)};X^{(1)}_{\cdot j}\bigr)
+
\ell\bigl(\hat\theta_j^{(2)};X^{(2)}_{\cdot j}\bigr)
-
\ell\bigl(\hat\theta_j^{(0)};X^{(0)}_{\cdot j}\bigr)
\Bigr),
\end{equation}
where \(X^{(0)}_{\cdot j}\) is the pooled sample and \(\hat\theta_j^{(0)}\), \(\hat\theta_j^{(1)}\), and \(\hat\theta_j^{(2)}\) are the maximum-likelihood estimates (MLEs) under the null and the two alternatives, respectively. Negative numerical values of \(\Lambda_j\), which may arise from finite-precision optimization, are truncated at zero. Under standard regularity conditions, Wilks' theorem yields the asymptotic null approximation \(\Lambda_j\sim\chi^2_{\nu}\), with \(\nu=2\) in both models considered below, since each univariate specification has two free parameters. The coordinate-wise \(p\)-value is therefore \(p_j=1-F_{\chi^2_2}(\Lambda_j)\). A change is declared if \(\min_j p_j<\alpha\) with \(\alpha=0.05\).

For Gaussian data, each coordinate is modeled as \(\mathcal{N}(\mu,\sigma^2)\). The MLEs admit the closed-form expressions \(\hat\mu=\bar x\) and \(\hat\sigma^2=n^{-1}\sum_{i=1}^{n}(x_i-\bar x)^2\), and the maximized log-likelihood reduces to
\begin{equation}
\ell(\hat\mu,\hat\sigma^2)
=
-\frac{n}{2}\bigl(\log(2\pi\hat\sigma^2)+1\bigr).
\end{equation}
The resulting two-parameter LR test therefore assesses equality of both mean and variance on each coordinate.

For Pareto data, each coordinate is modeled by the two-parameter family induced by the multivariate Pareto construction used in the simulations, with log-likelihood
\begin{equation}
\ell(\alpha,\beta)
=
\sum_{i=1}^{n}
\Bigl[
\log\beta
-
\bigl(\alpha^{-1}+1\bigr)
\log\bigl(1+\alpha\beta x_i\bigr)
\Bigr]
\end{equation}
for $\alpha,\beta>0$. The MLEs \((\hat\alpha,\hat\beta)\) are obtained numerically by maximizing \(\ell\) in the logarithmic parameterization \((\log\alpha,\log\beta)\). 
The LR statistic and the \(\chi^2_2\) \(p\)-value are then formed exactly as in the Gaussian case.

A permutation (conditional) variant of the same procedure was also considered. After pooling the two segments, \(B=199\) random reassignments into groups of size \(n\) are generated, the parametric LR statistics are recomputed on each permutation, and Monte Carlo \(p\)-values are formed in the usual way. Because the parametric statistic must be re-evaluated on every permutation, the computational cost scales as \(\Theta(B)\) times that of a single parametric LR evaluation. For the Gaussian model this factor already yields an approximately two-hundred-fold increase over the closed-form \(O(dn)\) parametric test; for the Pareto model the inflation is more severe, since each evaluation itself requires repeated numerical MLE fits. In view of this complexity, the permutation LR test is computationally prohibitive for the reported simulation study and is therefore not used.

\begin{table}[h!]
\centering
\caption{Likelihood-ratio and CHASM results for change scenarios for data from Multivariate Gaussian distribution.}
\label{tab:lr_chasm_change_results_gauss}
\begin{tabular}{lccc}
\hline
\textbf{id} & \textbf{$d$} & \textbf{LR-parametric} & \textbf{$ \mathcal{R} $} \\
\hline
1 & 2 & 1.00 & 0.99 \\
1 & 3 & 1.00 & 0.98 \\
1 & 5 & 1.00 & 0.96 \\
1 & 10 & 1.00 & 0.91 \\
1 & 25 & 1.00 & 0.84 \\
\hline
2 & 2 & 1.00 & 1.00 \\
2 & 3 & 1.00 & 1.00 \\
2 & 5 & 1.00 & 1.00 \\
2 & 10 & 1.00 & 1.00 \\
2 & 25 & 1.00 & 1.00 \\
\hline
3 & 2 & 1.00 & 0.95 \\
3 & 3 & 1.00 & 0.91 \\
3 & 5 & 1.00 & 0.93 \\
3 & 10 & 1.00 & 0.93 \\
3 & 25 & 1.00 & 0.93 \\
\hline
4 & 2 & 0.10 & 1.00 \\
4 & 3 & 0.16 & 1.00 \\
4 & 5 & 0.16 & 1.00 \\
4 & 10 & 0.44 & 1.00 \\
4 & 25 & 0.78 & 1.00 \\
\hline
5 & 2 & 0.12 & 1.00 \\
5 & 3 & 0.16 & 0.96 \\
5 & 5 & 0.14 & 0.97 \\
5 & 10 & 0.38 & 0.78 \\
5 & 25 & 0.82 & 0.43 \\
\hline
\end{tabular}
\end{table}

\begin{table}[h!]
\centering
\caption{Likelihood-ratio and CHASM results for no-change scenarios from Multivariate Gaussian distribution.}
\label{tab:lr_chasm_no_change_results_gauss}
\begin{tabular}{lccc}
\hline
\textbf{id} & \textbf{$d$} & \textbf{LR-parametric} & \textbf{$ \mathcal{R} $} \\
\hline
No change & 2 & 0.10 & 0.00 \\
No change & 3 & 0.22 & 0.00 \\
No change & 5 & 0.24 & 0.00 \\
No change & 10 & 0.44 & 0.00 \\
No change & 25 & 0.76 & 0.00 \\
\hline
\end{tabular}
\end{table}

\begin{table}[h!]
\centering
\caption{Likelihood-ratio and CHASM results for change scenarios for data from Multivariate Pareto distribution.}
\label{tab:lr_chasm_change_results_pareto}
\begin{tabular}{lccc}
\hline
\textbf{id} & \textbf{$d$} & \textbf{LR-parametric} & \textbf{$ \mathcal{R} $} \\
\hline
6 & 2 & 1.00 & 1.00 \\
6 & 3 & 1.00 & 1.00 \\
6 & 5 & 1.00 & 1.00 \\
6 & 10 & 1.00 & 1.00 \\
6 & 25 & 1.00 & 1.00 \\
\hline
7 & 2 & 1.00 & 1.00 \\
7 & 3 & 1.00 & 1.00 \\
7 & 5 & 1.00 & 1.00 \\
7 & 10 & 1.00 & 1.00 \\
7 & 25 & 1.00 & 1.00 \\
\hline
8 & 2 & 1.00 & 1.00 \\
8 & 3 & 1.00 & 1.00 \\
8 & 5 & 1.00 & 1.00 \\
8 & 10 & 1.00 & 1.00 \\
8 & 25 & 1.00 & 1.00 \\
\hline
9 & 2 & 1.00 & 1.00 \\
9 & 3 & 1.00 & 1.00 \\
9 & 5 & 1.00 & 1.00 \\
9 & 10 & 1.00 & 1.00 \\
9 & 25 & 1.00 & 1.00 \\
\hline
\end{tabular}
\end{table}

\begin{table}[h!]
\centering
\caption{Likelihood-ratio and CHASM results for no-change scenarios from Multivariate Pareto distribution.}
\label{tab:lr_chasm_no_change_results_pareto}
\begin{tabular}{lccc}
\hline
\textbf{id} & \textbf{$d$} & \textbf{LR-parametric} & \textbf{$ \mathcal{R} $} \\
\hline
No change & 2 & 0.08 & 0.04 \\
No change & 3 & 0.16 & 0.02 \\
No change & 5 & 0.26 & 0.02 \\
No change & 10 & 0.42 & 0.02 \\
No change & 25 & 0.56 & 0.02 \\
\hline
\end{tabular}
\end{table}

The same experiments as those described in Section~3 were performed for both data generated from multivariate Gaussian distributions and data generated from multivariate Pareto distributions. Tables~\ref{tab:lr_chasm_change_results_gauss}--\ref{tab:lr_chasm_no_change_results_pareto} report the ratio of simulations in which a change was classified as statistically significant using the parametric likelihood-ratio (LR) test and using CHASM, represented by the \textit{trustworthy factor} $\mathcal{R}$.

For the experiments based on the multivariate Gaussian distribution, the results presented in Tab.~\ref{tab:lr_chasm_change_results_gauss} show that both approaches achieve high accuracy for Scenarios~1--3. In Scenarios~4--5, which involve changes in the correlation structure, the LR-based approach exhibits substantially lower significance assessing performance. This behavior is expected because the considered LR test is based on an aggregated marginal formulation and therefore does not explicitly capture changes in the dependence structure between variables. Multivariate extensions of likelihood-ratio testing are available and could, in principle, address this limitation. However, such formulations are considerably more computationally demanding and typically require explicit modeling of the joint multivariate distribution. Furthermore, their statistical validity and robustness may depend strongly on the assumed parametric model. This constitutes an important limitation when the underlying data-generating distribution is unknown or does not conform well to the assumed model.

A more important distinction between the two approaches becomes apparent when considering the no-change experiments. As shown in Tab.~\ref{tab:lr_chasm_no_change_results_gauss}, the LR-based approach frequently classifies the data as exhibiting a statistically significant change despite the fact that both samples originate from the same distribution. The frequency of such false detections becomes particularly pronounced as the dimensionality $d$ increases. This indicates that, although the LR-based approach can achieve high sensitivity under scenarios with genuine changes, its reliability in distinguishing actual changes from random fluctuations may deteriorate in higher-dimensional settings. In contrast, this behavior is not observed to the same extent for CHASM, which maintains almost zero rate of false detections in the no-change scenario.

The experiments performed using multivariate Pareto distributions lead to similar conclusions. For scenarios containing an actual distributional change, both the LR-based approach and CHASM correctly assess the changes as significant, as shown in Tab.~\ref{tab:lr_chasm_change_results_pareto}. Thus, the ability of CHASM to assess the significance of changes is not limited to the Gaussian setting and remains effective for a substantially different, non-Gaussian distribution. However, the difference between the methods becomes apparent again in the no-change setting. As reported in Tab.~\ref{tab:lr_chasm_no_change_results_pareto}, the LR-based approach produces considerably more false detections than CHASM. Hence, the results obtained for both distribution families consistently indicate that high sensitivity alone is not sufficient to characterize the quality of a change-points authorization procedure. Controlling the rate of false authorization in the absence of an actual change is equally important, particularly when the method is intended for continuous monitoring of real-world data.

It should be emphasized that the LR-based approach considered in these experiments represents a particularly favorable case for parametric statistical testing. Specifically, the parametric test is evaluated under the assumption that the underlying distribution is known. Consequently, the LR procedure has access to information about the data-generating process that is generally unavailable in practical applications. This assumption substantially simplifies the testing problem and provides the LR-based approach with an advantage in controlled simulation experiments. In real-world applications, however, the underlying distribution is typically unknown and may change over time. Selecting an appropriate parametric model and verifying its validity therefore constitutes an additional source of uncertainty that is not present in the considered CHASM formulation.

CHASM follows a fundamentally different, nonparametric principle. It does not require the analyst to specify the distributional family from which the observations originate and does not rely on a predefined parametric model of the data. This makes CHASM particularly suitable for authorization of change-points for applications in which the distribution of the monitored process is unknown, difficult to model, or potentially misspecified by standard parametric assumptions.

Importantly, the absence of distributional assumptions does not imply that CHASM is insensitive to the structure of the data. Instead, the different CHASM criteria quantify complementary characteristics of the empirical distributions, allowing changes in location, scale, dependence, and more general distributional properties to be investigated.

A further practical consideration concerns computational complexity. Although multivariate parametric LR tests can in principle be constructed to account for changes in the joint distribution, because their computational cost increases substantially when moving from the considered marginal formulation to a fully multivariate implementation. 

The resulting execution time limits the practical applicability of such approaches, particularly in settings where statistical testing must be repeatedly performed over many windows or on high-dimensional data. This computational constraint is especially relevant for online or near-real-time change detection, where the testing procedure must be executed repeatedly as new observations become available.

Overall, the experiments demonstrate that the comparison between LR-based testing and CHASM should not be reduced to the probability of detecting a known change. Both approaches can achieve excellent sensitivity under favorable conditions. The more relevant distinction concerns their behavior in the absence of a change, their assumptions regarding the underlying distribution, and their computational requirements. The results indicate that CHASM provides a favorable compromise between sensitivity to genuine distributional changes, robustness against false detections, and applicability to data for which the underlying distribution cannot be reliably specified in advance. In this sense, the experiments support the use of CHASM as a general-purpose, nonparametric framework for assessing the statistical significance of changes in heterogeneous and potentially high-dimensional data.

\section{Window size for two-sample tests}
\label{sec:window_size}

An important aspect of methods that calculate statistical measures within selected windows is the appropriate selection of the window length. The window size directly determines the amount of data used to estimate the underlying distributions and, consequently, affects the reliability of the estimated test statistics. Short windows provide faster calculations; however, they contain fewer observations, which may result in high-variance estimates and reduced statistical power.  Therefore, the choice of the window length represents an important design parameter that should be explicitly investigated when applying CHASM to a particular data.

\begin{figure*}
    \centering
    \includegraphics[width=\linewidth]{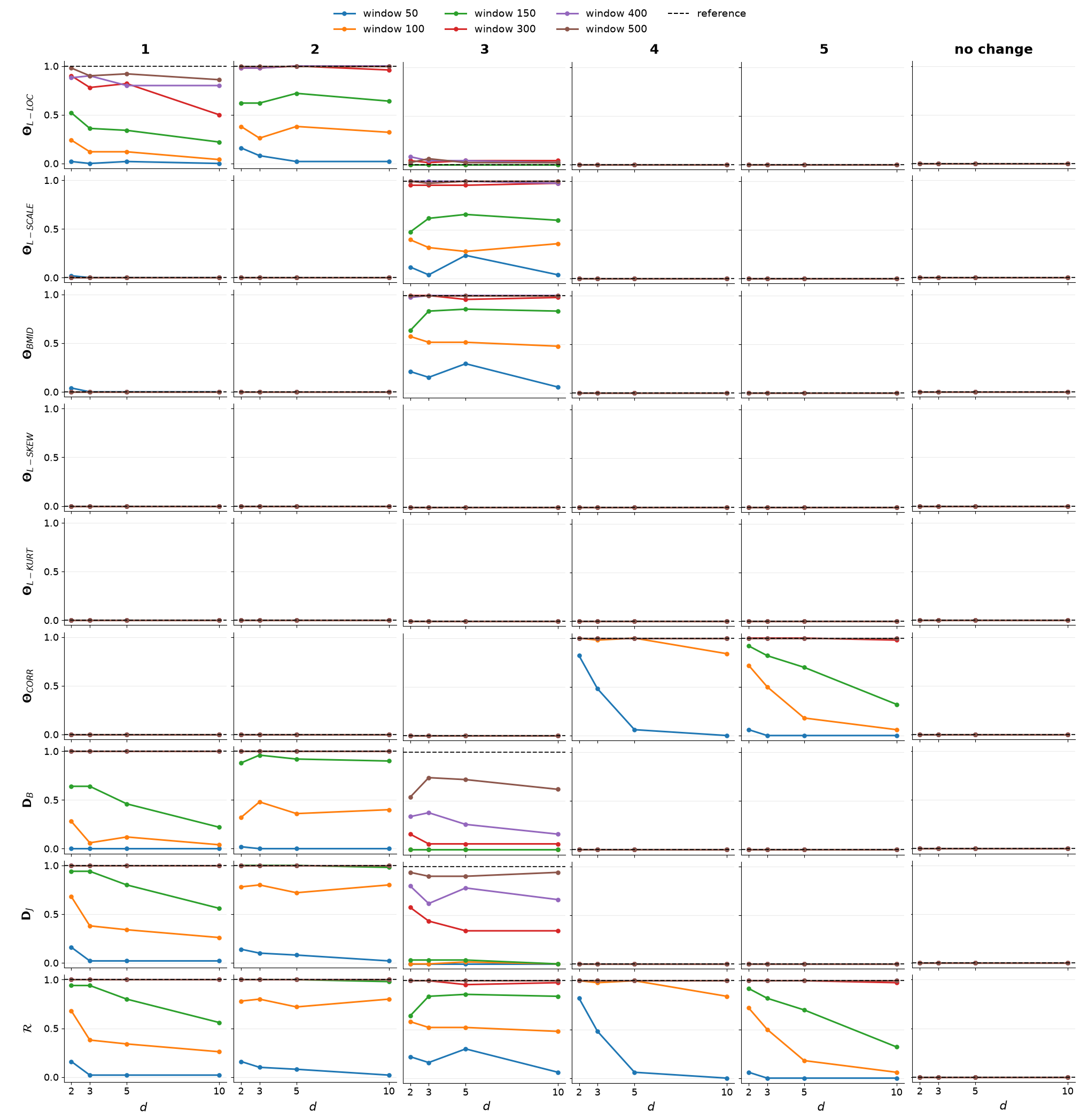}
    \caption{Evaluation of the impact of the window length $\omega$ on the quality of statistical significance testing based on individual criteria, depending on the dimension $d$ for different data-generation scenarios described in Section~3.1.}
    \label{fig:window_size}
\end{figure*}

In this appendix, we investigate the influence of the window length on the quality of statistical significance assessment of distributional changes by criterions used in CHASM. In particular, the purpose of this experiment is not to identify a universally optimal window length, but rather to demonstrate how the quality of statistical testing associated with a given set of criteria can be systematically evaluated. This is particularly important in the context of CHASM, where different criteria quantify different aspects of a distributional change. Consequently, window size that give us a satisfactory testing performance for one test statistic does not necessarily correspond to the same for another statistic. The appropriate window size for calculations of the statistic may depend on the properties of the statistic, the dimensionality of the data, and the amount of available data. These aspects should therefore be considered when selecting criteria and when designing a specific application of CHASM.

To investigate this effect, we perform simulations analogous to those described in Section~3.1 for several window lengths. More specifically, for each experiment listed in Table~\ref{tab:experiments_gauss}, we conduct 100 Monte Carlo simulations of the random vectors described by Eq.~\eqref{eg:gaussian_one_cp}. We consider data dimensions
$d \in {2,3,5,10,25}$ and window lengths
$\omega \in {50,100,150,300,400,500}$. For each simulation, the total number of observations is set to $N=2\omega$, with the reference change-point located at $n^*=\omega$. Thus, the two samples used by the two-sample tests contain $\omega$ observations each. In addition to the scenarios involving an actual distributional change, we evaluate the empirical false-positive behavior of the considered criteria using homogeneous data, for which no change is present.

For each simulation scenario (Scenarios 1--5 and the no-change scenario) and each dimensionality
$d \in \{2,3,5,10\}$, we compute the ratio of Monte Carlo simulations in which the change point is classified as statistically significant to the total number of Monte Carlo simulations. This quantity provides an empirical estimate of the probability that a given criterion detects the considered change under a particular combination of window length, dimensionality, and data-generating scenario. The results are presented in Fig.~\ref{fig:window_size}.

The rows of Fig.~\ref{fig:window_size} correspond to the individual statistical criteria described in Section~\ref{sec:statistics}, together with the \textit{trustworthy factor} $\mathcal{R}$. To facilitate visual interpretation, each plot includes a reference level (black dashed line) representing the expected detection rate under the corresponding scenario. The reference level is equal to 1 when the considered criterion is expected to detect the change in all simulations, and equal to 0 when no statistically significant change is expected.

The reference levels are defined according to the intended sensitivity of the individual criteria to the characteristics of the simulated changes. In Scenarios 1 and 2, the change concerns the location of the distribution. Therefore, criteria designed to detect changes in the mean or more generally changes in the distribution, namely $\Theta_{L-LOC}$, $D_B$, $D_J$, and $\mathcal{R}$, are expected to indicate a statistically significant change. In Scenario 3, the scale of the distribution changes. Accordingly, the criteria associated with scale changes and general distributional changes, i.e., $\Theta_{L-SCALE}$, $\Theta_{BMID}$, $D_B$, $D_J$, and $\mathcal{R}$, are expected to detect the change. In Scenarios 4 and 5, the change concerns the dependence structure, specifically the correlation between the variables. A statistically significant change is therefore expected only for the criteria $\Theta_{CORR}$ and $\mathcal{R}$, for which the reference level is set to 1. In contrast, no change is expected for $D_B$ and $D_J$, whose values are computed from the marginal distributions and are consequently unaffected by changes in the correlation structure; their reference level is therefore set to 0. Finally, for the no-change scenario, the reference level is set to 0 for all criteria, as no statistically significant change is expected.

As can be observed, window lengths ranging from 50 to 150 exhibit a pronounced divergence in performance for many considered scenarios. In general, the detection efficiency (where efficiency for reference=1 is defined as the ratio of the number of simulations in which a statistically significant change was detected to the total number of simulations) increases with the window length. e.g., for Scenario 2 it increases up to approximately window size =150. Above this value, the considered divergence measures achieve an efficiency close to 1. For shorter window lengths, the detection efficiency is relatively low and decreases with increasing dimensionality $d$.

The most pronounced variation can be observed for $\Theta_{L-LOC}$ in Scenario 1. In this case, the change in location is relatively small, affecting only a single dimension, which indicates that a longer observation window is required for reliable detection. In Scenario 2, the change is more pronounced, as it affects all dimensions, and the results obtained for window lengths between 300 and 500 are very similar.

Scenario 3, corresponding to a scale change, demonstrates that the optimal window length may depend on both the characteristics of the data and the employed test statistic. In this scenario, the $\Theta_{L-SCALE}$ and $\Theta_{BMID}$ criteria achieve an efficiency close to 1 for window lengths of at least 300. The distribution-based criterion $D_J$ also provides relatively good performance in detecting significant changes for a window length of 500, whereas $D_B$ exhibits low efficiency for all considered window lengths.

The correlation-change scenarios further demonstrate the influence of the window length on detection performance. Short windows (100--150) are sufficient to detect substantial correlation changes, as illustrated by Scenario 4, but are insufficient for detecting more subtle changes, as demonstrated in Scenario 5. For window lengths greater than 150, however, the detection efficiency is close to 1 in both scenarios.

Finally, in the 'no change' scenario, the results remain consistent regardless of the window length. In all cases, the number of detections indicating a statistically significant change is close to zero, which is consistent with the expected behavior of the considered methods under the null scenario.

\section{Additional results for contaminated multivariate Gaussian model}
\label{app:cont_gauss}

\begin{figure*}
    \centering
    \includegraphics[width=\linewidth]{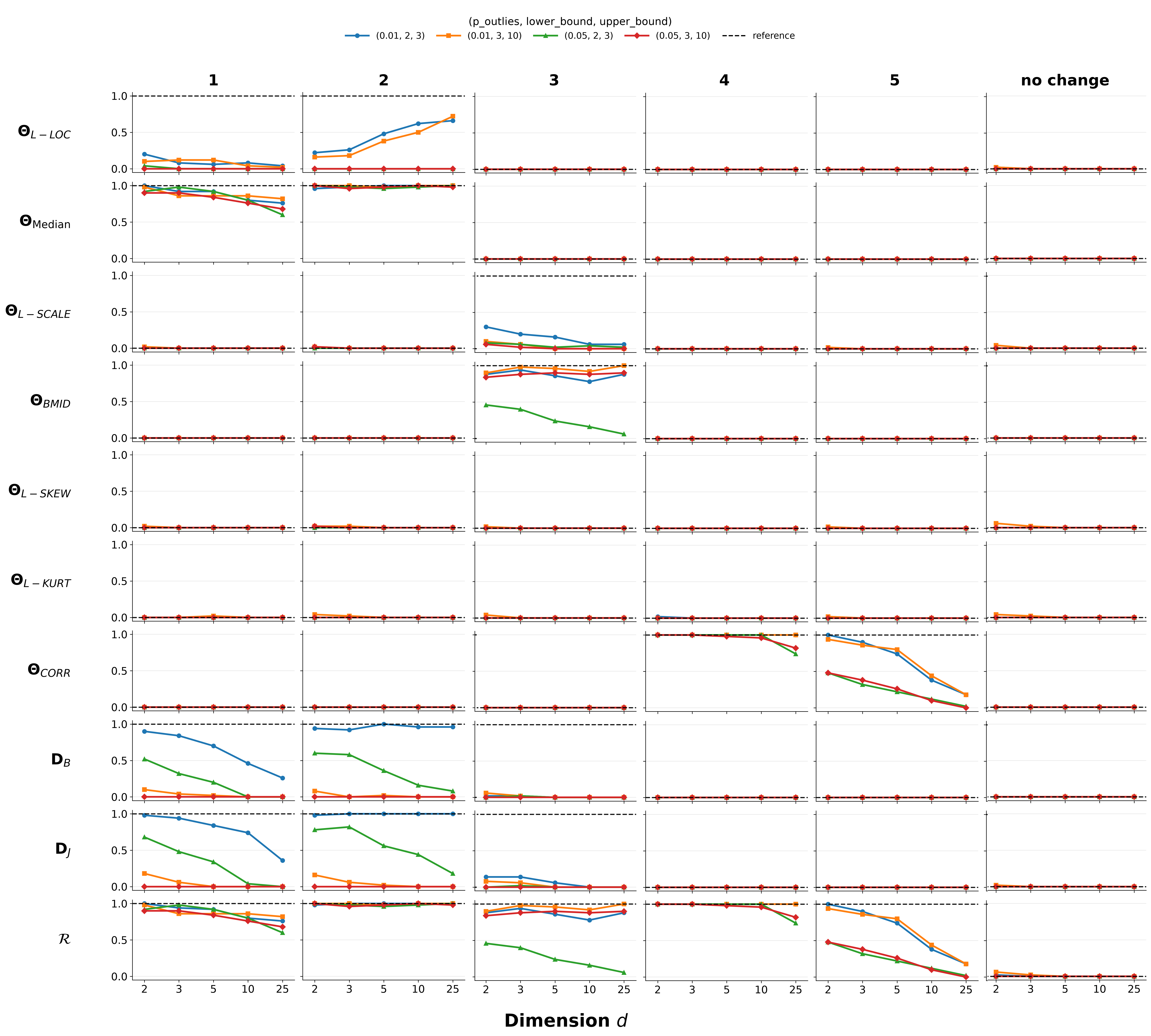}
    \caption{Evaluation of the impact of the contamination level on the quality of statistical significance testing based on individual criteria, depending on the dimension $d$ for different data-generation scenarios described in Section~3.1 and $\mathbf{X}_t$ modelled by Eq.~\eqref{eq:impulse_contam}.}
    \label{fig:contaminated}
\end{figure*}

In this section, we consider \(d\)-dimensional Gaussian vectors that follow the model in~\eqref{eq:assumption_multiple_cp} and are additionally corrupted by additive, sign-preserving impulse outliers.
For \(\mathbf{X}_1,\mathbf{X}_2,\ldots,\mathbf{X}_N\) as in~\eqref{eq:assumption_multiple_cp}, we write
\(\mathbf{X}_{i:j}=\{\mathbf{X}_i,\mathbf{X}_{i+1},\ldots,\mathbf{X}_j\}\)
for \(i,j\in\mathbb{N}\) with \(i<j\), and let \(X_{i:j}^{k}\) denote the sample of \(k\)th coordinates of \(\mathbf{X}_{i:j}\), \(k=1,\ldots,d\).
The corresponding contaminated observations \(\mathbf{Y}_{i:j}\) are defined, for each \(t=i,\ldots,j\) and each coordinate \(k=1,\ldots,d\), by
\begin{equation}
Y_{t}^{k}
=
X_{t}^{k}
+
\operatorname{sign}\!\bigl(X_{t}^{k}\bigr)\,
K_{t}^{k}\,
W_{t}^{k},
\label{eq:impulse_contam}
\end{equation}
where
\begin{equation}
\operatorname{sign}(x)
=
\begin{cases}
-1, & x<0,\\
\phantom{-}0, & x=0,\\
\phantom{-}1, & x>0,
\end{cases}
\label{eq:sign_function}
\end{equation}
and the sequences \(\{K_{t}^{k}\}\) and \(\{W_{t}^{k}\}\) are independent of \(\{\mathbf{X}_{t}\}\) and mutually independent across both time and coordinates, with
\begin{equation}
K_{t}^{k}\sim\mathcal{U}(a_{k},b_{k}),
\qquad
W_{t}^{k}\sim\mathrm{Bernoulli}(p_{out}),
\label{eq:impulse_KW}
\end{equation}
\(b_{k}>a_{k}>0\), and \(p_{out}\in[0,1]\).
In vector form,~\eqref{eq:impulse_contam} is equivalent to
\begin{equation}
\mathbf{Y}_{t}
=
\mathbf{X}_{t}
+
\operatorname{sign}(\mathbf{X}_{t})\odot\mathbf{K}_{t}\odot\mathbf{W}_{t},
\label{eq:impulse_vector}
\end{equation}
where \(\operatorname{sign}(\cdot)\) acts component-wise and \(\odot\) denotes the Hadamard product.
The Bernoulli parameter \(p_{out}\) controls the frequency of impulses, whereas \(a_{k}\) and \(b_{k}\) control their amplitude on coordinate \(k\).
To keep the amplitude comparable across coordinates of different scale, the bounds \((a_{k},b_{k})\) are tied to \(X_{i:j}^{k}\).
Let \(Q_{i:j}^{k}\) denote the empirical \(0.95\)-quantile of the uncontaminated absolute sample \(\bigl\{|X_{t}^{k}|:t=i,\ldots,j\bigr\}\).
We set
\begin{equation}
a_{k}=c_{\mathrm{L}}\,Q_{i:j}^{k},
\qquad
b_{k}=c_{\mathrm{U}}\,Q_{i:j}^{k},
\label{eq:impulse_quantile}
\end{equation}
with \(c_{\mathrm{U}}>c_{\mathrm{L}}>0\).
Contamination is applied independently to each of the two segments entering the test.
In the experiments we take \(p_{out}\in\{0.01,0.05,0.10\}\) and \((c_{\mathrm{L}},c_{\mathrm{U}})\in\{(2,3),(3,10)\}\).

For each simulation scenario (Scenarios 1--5 and the no-change scenario) and each dimensionality $d \in {2,3,5,10, 25}$, we repeat the simulations described in Section~3.1, replacing $\mathbf{X}_t$ drawn from a multivariate Gaussian distribution with data generated according to Eq.~\eqref{eq:impulse_contam}. Next, we compute the ratio of Monte Carlo simulations in which the change point is classified as statistically significant to the total number of Monte Carlo simulations. This quantity provides an empirical estimate of the probability that a given criterion detects the considered change under a particular combination of $p_{out}, c_{\mathrm{L}},c_{\mathrm{U}}$, $d$, and data-generating scenario. The results are presented in Fig.~\ref{fig:contaminated}. 

As described in \ref{sec:window_size} - to facilitate visual interpretation, each plot includes a reference level (black dashed line) corresponding to the expected detection rate for the given scenario. The reference level is set to 1 when a criterion is expected to detect the change in all simulation runs, and to 0 when no statistically significant change is expected.

The reference levels are determined based on the sensitivity of each criterion to the specific characteristics of the simulated changes. In Scenarios 1 and 2, the change affects the location of the distribution. Hence, criteria targeting changes in the mean or, more generally, changes in the underlying distribution (namely, $\Theta_{L-LOC}$, $D_B$, $D_J$, and $\mathcal{R}$) are expected to signal a statistically significant change. In Scenario 3 the scale of the distribution is altered. Consequently, criteria designed to detect changes in scale or more general distributional changes (namely $\Theta_{L-SCALE}$, $\Theta_{BMID}$, $D_B$, $D_J$, and $\mathcal{R}$) are expected to detect the change. In Scenarios 4 and 5, the change concerns the dependence structure, specifically the correlation between the variables. A statistically significant change is therefore expected only for the criteria $\Theta_{CORR}$ and $\mathcal{R}$, for which the reference level is set to 1. In contrast, no change is expected for $D_B$ and $D_J$, whose values are computed from the marginal distributions and are consequently unaffected by changes in the correlation structure; their reference level is therefore set to 0. Finally, for the no-change scenario, the reference level is set to 0 for all criteria, as no statistically significant change is expected.

{In addition to the robust estimator introduced previously, based on the difference in robust location estimates, $\Theta_{L\text{-}LOC}$, an analogous criterion based on the difference in medians, $\Theta_{\mathrm{MEDIAN}}$, is also considered in this section. This criterion is included to demonstrate that the choice of an appropriate estimator, insensitive to irrelevant information (in this case, non-Gaussian impulses embedded in the noise), is important when assessing the significance of regime changes.}

{As shown in Scenarios 1 and 2, only $\Theta_{\mathrm{MEDIAN}}$ provides effective change detection regardless of the intensity of the additional impulses. The divergence-based measures perform adequately only in the presence of isolated impulses with relatively small amplitudes. For changes in scale (Scenario 3), the robust scale-difference estimators, $\Theta_{L\text{-}SCALE}$ and $\Theta_{\mathrm{BMID}}$, also exhibit different levels of effectiveness. In contrast, the divergence-based estimators, $D_B$ and $D_J$, fail to indicate a significant change in cases where such a change is expected. These results indicate that, for highly contaminated data, the behavior of the adopted estimators should be carefully investigated. Importantly, robust estimators based on trimmed statistics may benefit from increasing the length of the observation window used for testing, as discussed in Appendix~C. Therefore, before applying a given criterion to a specific real-world use case, it is recommended to perform experiments analogous to those presented in Appendix~C in order to assess its behavior under conditions representative of the target application.}
{Scenarios 4--5 further demonstrate that the performance of correlation-change detection based on $\Theta_{\mathrm{CORR}}$ decreases substantially as the dimensionality $d$ increases. In the case of small correlation changes (Scenario 5) and frequent impulses ($p=0.05$), the proportion of simulations in which a statistically significant correlation change is detected falls below $0.5$. To mitigate this effect, a more robust estimator or longer observation windows, as investigated in Appendix~C, should be considered for such settings.}

{Finally, in the absence of a regime change, all considered criteria correctly indicate no significant change in nearly all simulations. This demonstrates that, despite the differences in detection performance observed under contaminated conditions, the considered estimators exhibit a low false-positive rate when the underlying regime remains unchanged.
}

\section{Additional results for multidimensional industrial data}
\label{app:simulated_data1}
Tab.~\ref{tab1} -- Tab.~\ref{tab3} present the remaining results regarding the study of the explainability of regime changes, which, based on selected changes, were identified by CHASM as statistically significant. The tables are supplementary to Section \ref{sec:real_data} of the main part of the article.

\begin{table*}[b!]
\centering
\caption{Explainability analysis for change point $n_2=3533$  for Dataset 1.}
\label{tab1}
\begin{tabular}{l|l|l|l|l|l|l|l|l}
\hline
\textbf{cp}             & \textbf{variable} & \textbf{$\Theta_{L-SCALE}$}             & \textbf{$\Theta_{L-LOC}$}              & \textbf{$\Theta_{BMID}$}        & \textbf{$\Theta_{L-SKEW}$}          & \textbf{$\Theta_{L-KURT}$}          & \textbf{$D_B$}  & \textbf{$D_J$}               \\ \hline
                        & v1                &                               &                               &                               &                               &                               & \cellcolor[HTML]{FFEB84}0.631  & \cellcolor[HTML]{F5E883}0.395        \\
                        & v2                &                               &                               &                               &                               &                               & \cellcolor[HTML]{FFEB84}0.697  & \cellcolor[HTML]{FFEB84}0.444        \\
                        & v3                & \cellcolor[HTML]{99CD7E}0.155 &                               & \cellcolor[HTML]{C6DA80}0.270 &                               &                               & \cellcolor[HTML]{D6DF81}0.313  & \cellcolor[HTML]{C9DB80}0.279        \\
                        & v4                & \cellcolor[HTML]{9DCF7E}0.165 & \cellcolor[HTML]{75C37C}0.059 & \cellcolor[HTML]{95CC7D}0.144 & \cellcolor[HTML]{FFEB84}0.521 &                               & \cellcolor[HTML]{FFEB84}0.752  & \cellcolor[HTML]{FFEB84}9.051        \\
                        & v5                &                               &                               &                               &                               &                               &                                &                                      \\
                        & v6                &                               &                               &                               &                               &                               &                                &                                      \\
                        & v7                &                               &                               &                               &                               &                               &                                &                                      \\
                        & v8                &                               &                               &                               &                               &                               &                                &                                      \\
                        & v9                & \cellcolor[HTML]{FFEB84}0.448 & \cellcolor[HTML]{75C37C}0.058 & \cellcolor[HTML]{FFEB84}0.843 & \cellcolor[HTML]{FFEB84}1.095 &                               &                                & \cellcolor[HTML]{F8696B}70431307.960 \\
                        & v10               &                               &                               &                               &                               &                               &                                &                                      \\
                        & v11               & \cellcolor[HTML]{FFEB84}0.781 & \cellcolor[HTML]{7BC57C}0.075 &                               &                               &                               & \cellcolor[HTML]{C6DA80}0.272  & \cellcolor[HTML]{A4D17E}0.184        \\
                        & v12               &                               &                               &                               &                               &                               &                                &                                      \\
                        & v13               &                               &                               &                               &                               &                               &                                &                                      \\
                        & v14               &                               &                               &                               &                               &                               &                                &                                      \\
                        & v15               & \cellcolor[HTML]{86C87D}0.103 &                               & \cellcolor[HTML]{BDD880}0.248 &                               &                               & \cellcolor[HTML]{FFEB84}14.715 & \cellcolor[HTML]{FFEB84}6.221        \\
                        & v16               & \cellcolor[HTML]{FFEB84}0.493 & \cellcolor[HTML]{63BE7B}0.011 & \cellcolor[HTML]{FFEB84}0.546 &                               & \cellcolor[HTML]{A2D07E}0.178 &                                &                                      \\
                        & v17               & \cellcolor[HTML]{FFEB84}0.660 &                               & \cellcolor[HTML]{FFEB84}0.867 &                               & \cellcolor[HTML]{F3E783}0.388 & \cellcolor[HTML]{FFEB84}0.584  & \cellcolor[HTML]{FFEB84}1.340        \\
\multirow{-18}{*}{3533} & v18               &                               &                               & \cellcolor[HTML]{B0D47F}0.213 & \cellcolor[HTML]{9FCF7E}0.169 &                               &                                &                                      \\ \hline
\end{tabular}
\end{table*}

\begin{table*}[t!]
\centering
\caption{Explainability analysis for change point $n_3=3931$  for Dataset 1.}
\begin{tabular}{l|l|l|l|l|l|l|l|l}
\hline
\textbf{cp}             & \textbf{variable} & \textbf{$\Theta_{L-SCALE}$}             & \textbf{$\Theta_{L-LOC}$}              & \textbf{$\Theta_{BMID}$}        & \textbf{$\Theta_{L-SKEW}$}          & \textbf{$\Theta_{L-KURT}$}          & \textbf{$D_B$}  & \textbf{$D_J$}        \\ \hline
                        & v1                &                               & \cellcolor[HTML]{82C77C}0.081 &                               & \cellcolor[HTML]{F3E783}0.367 &                               &                                &                               \\
                        & v2                & \cellcolor[HTML]{7EC67C}0.071 & \cellcolor[HTML]{90CB7D}0.115 &                               &                               &                               & \cellcolor[HTML]{FFE984}0.639  & \cellcolor[HTML]{FFEA84}0.562 \\
                        & v3                & \cellcolor[HTML]{A3D07E}0.164 & \cellcolor[HTML]{9BCE7E}0.142 &                               &                               &                               & \cellcolor[HTML]{FECA7E}3.812  & \cellcolor[HTML]{FECC7E}3.629 \\
                        & v4                &                               &                               &                               & \cellcolor[HTML]{FFE683}0.944 &                               &                                &                               \\
                        & v5                &                               &                               &                               & \cellcolor[HTML]{BFD880}0.235 &                               &                                &                               \\
                        & v6                &                               &                               &                               & \cellcolor[HTML]{FFEB84}0.429 &                               &                                &                               \\
                        & v7                &                               &                               &                               & \cellcolor[HTML]{CDDC81}0.270 &                               &                                &                               \\
                        & v8                &                               &                               &                               &                               &                               &                                &                               \\
                        & v9                &                               &                               & \cellcolor[HTML]{FFE583}1.038 &                               &                               &                                &                               \\
                        & v10               &                               & \cellcolor[HTML]{A3D07E}0.163 &                               &                               &                               & \cellcolor[HTML]{F8696B}13.766 &                               \\
                        & v11               &                               &                               &                               &                               &                               & \cellcolor[HTML]{FFE884}0.766  & \cellcolor[HTML]{EEE683}0.352 \\
                        & v12               &                               &                               &                               &                               &                               &                                &                               \\
                        & v13               & \cellcolor[HTML]{8AC97D}0.100 &                               &                               &                               & \cellcolor[HTML]{74C37C}0.045 &                                &                               \\
                        & v14               &                               &                               &                               &                               &                               &                                &                               \\
                        & v15               &                               &                               & \cellcolor[HTML]{FFE483}1.209 &                               &                               & \cellcolor[HTML]{FFE383}1.252  & \cellcolor[HTML]{FFE884}0.720 \\
                        & v16               &                               &                               &                               &                               & \cellcolor[HTML]{FFEB84}0.422 &                                &                               \\
                        & v17               &                               &                               &                               &                               &                               &                                &                               \\
\multirow{-18}{*}{3931} & v18               &                               &                               &                               &                               & \cellcolor[HTML]{63BE7B}0.001 &                                &                               \\ \hline
\end{tabular}
\end{table*}

\begin{table*}[t!]
\centering
\caption{Explainability analysis for change point $n_4=4928$  for Dataset 1.}
\label{tab3}
\begin{tabular}{l|l|l|l|l|l|l|l|l}
\hline
\textbf{cp}             & \textbf{variable} & \textbf{$\Theta_{L-SCALE}$}             & \textbf{$\Theta_{L-LOC}$}              & \textbf{$\Theta_{BMID}$}        & \textbf{$\Theta_{L-SKEW}$}          & \textbf{$\Theta_{L-KURT}$}          & \textbf{$D_B$}  & \textbf{$D_J$}        \\ \hline
                        & v1                & \cellcolor[HTML]{FFEA84}0.764 & \cellcolor[HTML]{DBE081}0.530 & \cellcolor[HTML]{8ECA7D}0.202 &                               & \cellcolor[HTML]{6BC07B}0.056 &                                &                               \\
                        & v2                & \cellcolor[HTML]{FFEA84}0.800 & \cellcolor[HTML]{CDDC81}0.467 & \cellcolor[HTML]{8ECA7D}0.205 &                               & \cellcolor[HTML]{6AC07B}0.053 &                                &                               \\
                        & v3                & \cellcolor[HTML]{FFEB84}0.678 & \cellcolor[HTML]{FAE983}0.661 & \cellcolor[HTML]{8BC97D}0.192 &                               &                               &                                &                               \\
                        & v4                & \cellcolor[HTML]{63BE7B}0.020 & \cellcolor[HTML]{85C87D}0.167 &                               &                               &                               & \cellcolor[HTML]{FCA577}6.073  & \cellcolor[HTML]{FECB7E}3.185 \\
                        & v5                &                               &                               &                               &                               &                               &                                &                               \\
                        & v6                &                               &                               &                               &                               &                               &                                &                               \\
                        & v7                &                               &                               &                               &                               &                               &                                &                               \\
                        & v8                &                               &                               &                               &                               &                               &                                &                               \\
                        & v9                & \cellcolor[HTML]{B6D57F}0.370 &                               & \cellcolor[HTML]{FFE784}0.987 &                               &                               &                                &                               \\
                        & v10               &                               &                               &                               &                               &                               &                                & \cellcolor[HTML]{FFE984}0.907 \\
                        & v11               & \cellcolor[HTML]{F0E683}0.617 &                               &                               &                               &                               & \cellcolor[HTML]{F8696B}10.608 & \cellcolor[HTML]{FDBF7C}4.084 \\
                        & v12               &                               &                               &                               &                               &                               &                                &                               \\
                        & v13               &                               &                               &                               &                               &                               &                                &                               \\
                        & v14               & \cellcolor[HTML]{67BF7B}0.039 &                               &                               &                               &                               &                                &                               \\
                        & v15               & \cellcolor[HTML]{C9DB80}0.453 &                               & \cellcolor[HTML]{FFEB84}0.696 &                               &                               & \cellcolor[HTML]{FFE984}0.876  & \cellcolor[HTML]{FFEB84}0.747 \\
                        & v16               &                               &                               &                               &                               &                               & \cellcolor[HTML]{FB9C75}6.759  & \cellcolor[HTML]{FCA577}6.080 \\
                        & v17               &                               &                               &                               & \cellcolor[HTML]{A8D17E}0.311 &                               & \cellcolor[HTML]{FECF7F}2.888  & \cellcolor[HTML]{FFDE82}1.733 \\
\multirow{-18}{*}{4928} & v18               &                               &                               &                               &                               &                               &                                &                               \\ \hline
\end{tabular}
\end{table*}

\end{document}